\documentclass[prb,aps,twocolumn,eqsecnum,amsmath,amssymb,showpacs]{revtex4}

\usepackage{dcolumn}% Align table columns on decimal point
\usepackage{bm}% bold math
\usepackage{graphicx}

\begin{document}

\title{       One-dimensional orbital fluctuations and 
              the exotic magnetic properties of YVO$_3$ }

\author{      Andrzej M. Ole\'{s} }
\affiliation{ Max-Planck-Institut f\"ur Festk\"orperforschung, 
	      Heisenbergstrasse 1, D-70569 Stuttgart, Germany \\
              Marian Smoluchowski Institute of Physics, Jagellonian
              University, Reymonta 4, PL-30059 Krak\'ow, Poland }

\author{      Peter Horsch }
\affiliation{ Max-Planck-Institut f\"ur Festk\"orperforschung,
              Heisenbergstrasse 1, D-70569 Stuttgart, Germany }

\author{      Giniyat Khaliullin }
\affiliation{ Max-Planck-Institut f\"ur Festk\"orperforschung,
              Heisenbergstrasse 1, D-70569 Stuttgart, Germany } 

\date{21 December 2006}

\begin{abstract}
Starting from the Mott insulator picture for cubic vanadates, we derive
and investigate the model of superexchange interactions between V$^{3+}$
ions, with nearly degenerate $t_{2g}$ orbitals occupied by two electrons
each. The superexchange interactions are strongly frustrated and 
demonstrate a strong interrelation between possible types of magnetic 
and orbital order. We elucidate the prominent role played by fluctuations 
of $yz$ and $xz$ orbitals which generate ferromagnetic superexchange 
interactions even in the absence of Hund's exchange. In this limit we 
find orbital valence bond state which is replaced either by $C$-type 
antiferromagnetic order with weak $G$-type orbital order at increasing 
Hund's exchange, or instead by $G$-type antiferromagnetic order when the 
lattice distortions stabilize $C$-type orbital order. Both phases are 
observed in YVO$_3$ and we argue that a dimerized $C$-type 
antiferromagnetic phase with stronger and weaker FM bonds alternating 
along the $c$ axis may be stabilized by large spin-orbital entropy at 
finite temperature. This suggests a scenario which explains the origin of 
the exotic $C$-AF order observed in YVO$_3$ in the regime of intermediate 
temperatures and allows one to specify the necessary ingredients of a 
more complete future theory.\\
{\it [Published in: Phys. Rev. B {\bf 75}, 184434 (2007).]}
\end{abstract}


\maketitle

%%%%%%%%%%%%%%%%%%%%%%%%%%%%%%%%%%%%%%%%%%%%%%%%%%%%%%%%%%%%%%%%%%%%%%%%%
%%                         Orbital degrees of freedom
%%%%%%%%%%%%%%%%%%%%%%%%%%%%%%%%%%%%%%%%%%%%%%%%%%%%%%%%%%%%%%%%%%%%%%%%%
\section{ Orbital degrees of freedom }
\label{sec:orbi}

%%%%%%%%%%%%%%%%%%%%%%%%%%%%%%%%%%%%%%%%%%%%%%%%%%%%%%%%%%%%%%%%%%%%%%%%
%%                     orbital physics -- general       
%%%%%%%%%%%%%%%%%%%%%%%%%%%%%%%%%%%%%%%%%%%%%%%%%%%%%%%%%%%%%%%%%%%%%%%%
Transition metal oxides with the perovskite structure display a large
variety of properties such as high-temperature superconductivity and 
colossal magnetoresistance. Their magnetic properties are also quite
diverse, with antiferromagnetic (AF), disordered, or ferromagnetic (FM) 
phases in different doping regimes, being the subject of particularly 
active research in the last decade.\cite{Ima98,Mae04} Although certain 
universal principles can be formulated, these complex magnetic 
properties depend on the actual filling of $3d$ orbitals of transition 
metal ions, and have to be studied in detail for each family of 
compounds separately. Rich and complex behavior in doped systems is 
found as moving charges can dress by spin or orbital excitations.
\cite{Zaa93} The undoped compounds are somewhat simpler as their 
properties are dominated by large on-site Coulomb interactions 
$\propto U$, responsible for their Mott-Hubbard (or charge-transfer) 
insulating behavior, with the effective low-energy magnetic 
interactions of superexchange type. While such interactions are AF and 
nonfrustrated on a cubic lattice 
for nondegenerate orbitals, they have a very nontrivial structure when 
degenerate $3d$ orbitals are partly occupied, as pointed out by Kugel 
and Khomskii on the example of $e_g$ systems long ago.\cite{Kug82} 
In such cases the {\it orbital degrees of freedom\/} have to be 
considered on equal footing with electron spins,\cite{Kug82,Goo63} 
which leads to the so-called spin-orbital superexchange models,
\cite{Tok00,Ole01,Kha05,Ole05} describing the low-energy physics
and the partial sum rules in the optical spectroscopy.\cite{Kha04} 

%%%%%%%%%%%%%%%%%%%%%%%%%%%%%%%%%%%%%%%%%%%%%%%%%%%%%%%%%%%%%%%%%%%%%%%%
%%                 coupling of spins and orbital in eg 
%%%%%%%%%%%%%%%%%%%%%%%%%%%%%%%%%%%%%%%%%%%%%%%%%%%%%%%%%%%%%%%%%%%%%%%%
An intriguing feature of the spin-orbital models is the strong
frustration of the superexchange interactions on a cubic (perovskite)
lattice which was recognized as the origin of enhanced quantum effects
in transition metal oxides.\cite{Fei97} For purely electronic models
this frustration might even lead to the collapse of long range order in 
particular parameter regimes, but usually this does not happen and the
fluctuations are partly suppressed either by the order-out-of-disorder
mechanism,\cite{Kha97} or by the coupling to the lattice distortions
induced by the Jahn-Teller (JT) effect. In the cuprates and manganites
both the superexchange and the JT interactions support each other,
\cite{Fei99,Oka02} and such systems undergo usually structural 
transitions. Nevertheless, even below the structural transition the
spin and orbital degrees of freedom are coupled, leading to 
characteristic changes of the orbital order (OO) at magnetic transitions 
and to new composite spin-orbital excitations, when both spin and orbital 
excitation occurs simultaneously.\cite{Ole00} 

%%%%%%%%%%%%%%%%%%%%%%%%%%%%%%%%%%%%%%%%%%%%%%%%%%%%%%%%%%%%%%%%%%%%%%%%
%%                 orbitals in t2g: titanates and v2o3
%%%%%%%%%%%%%%%%%%%%%%%%%%%%%%%%%%%%%%%%%%%%%%%%%%%%%%%%%%%%%%%%%%%%%%%%
The importance of the orbital degrees of freedom was realized in the
theory of magnetism already in the seventies. Next to $e_g$ systems,
\cite{Kug73} model Hamiltonians with twofold degeneracy and diagonal 
hopping,\cite{Cyr75} and the realistic effecrive Hamiltonian for 
$t_{2g}$ electrons in V$_2$O$_3$ were 
studied.\cite{Cas78} Actually, the superexchange interactions for partly 
filled $t_{2g}$ orbitals are different and even more fascinating than 
those for $e_g$ systems. As realized first for the $d^1$ configuration 
in cubic titanates,\cite{Kha00} the quantum effects are here even 
stronger than in the $e_g$ systems (cuprates or manganites), 
as the JT coupling is weak and the orbitals may form the 
coherent orbital liquid state observed in a Mott insulator LaTiO$_3$. 
\cite{Kei00} As a result of this quantum behavior and common 
spin-orbital fluctuations, the classical Goodenough-Kanamori rules are 
violated in $t_{2g}$ systems in some cases.\cite{Ole06} 

The quantum effects are equally important in vanadium compounds with 
V$^{3+}$ ions in the $d^2$ configuration, realized in V$_2$O$_3$ and in 
cubic compounds: LaVO$_3$ and YVO$_3$. The metal-insulator transition in 
V$_2$O$_3$ is studied for quite a long time,\cite{Ima98} but more 
realistic superexchange models were introduced only after the 
experimental evidence of the OO which occurs below the 
magnetic transition.\cite{Bao97} The first spin-orbital model for 
V$_2$O$_3$ was assuming a picture of molecular bonds which saturated one 
$t_{2g}$ electron per V$^{3+}$ site and thus used $s=1/2$ spins.\cite{Cas78} However,
one decade ago it was realized that Hund's exchange $J_H$ is large,
\cite{Miz96} and the superexchange interactions couple instead $S=1$ 
spins of different V$^{3+}$ ions. A complete superexchange model with spin and orbital 
degrees of freedom in V$_2$O$_3$ was derived only a few years 
ago by Di Matteo, Perkins and Natoli.\cite{Dim02}

%%%%%%%%%%%%%%%%%%%%%%%%%%%%%%%%%%%%%%%%%%%%%%%%%%%%%%%%%%%%%%%%%%%%%%%%%
%%                    cubic vanadates: experimental 
%%%%%%%%%%%%%%%%%%%%%%%%%%%%%%%%%%%%%%%%%%%%%%%%%%%%%%%%%%%%%%%%%%%%%%%%%
Also in cubic vanadates evidence increases that the orbital degrees of
freedom couple to the magnetic order and play an important and highly 
nontrivial role. In LaVO$_3$ the $C$-type AF ($C$-AF) phase [with FM 
chains along $c$ axis which stagger within $ab$ planes] is stable below 
N\'eel temperature $T_N\simeq 143$ K, followed by a weak structural 
transition at $T_s\simeq 141$ K.\cite{Mah92,Goo95,Miy00,Ren03,Miy03,Miy06} 
Remarkably, the magnetic order parameter in the $C$-AF phase of LaVO$_3$ 
is strongly reduced to $\simeq 1.3\mu_B$.\cite{Goo95} As the spin quantum 
fluctuations are smaller than in the $G$-AF phase and are unlikely to 
decrease the order parameter by more than 6\% for $S=1$ spins,\cite{Rac02} 
the observed large reduction of the magnetic moments suggests that some 
other quantum effects which originate from the 
orbital degeneracy dominate in this phase of cubic vanadates.

The situation is very different and even more puzzling in YVO$_3$
\cite{Kaw94,Ren00,Bla01,Miy03,Miy06,Ree06} --- this compound has $G$-type 
AF order (staggered in all three directions, called below $G$-AF) at low 
temperatures $T<T_{N2}$, while the magnetic order changes in the first 
order magnetic transition at $T_{N2}=77$ K to $C$-AF structure which 
remains stable up to $T_{N1}\simeq 116$ K. The magnetic transition at 
$T_{N2}$ is particularly surprising as the staggered moments change 
their direction from approximately parallel to the $c$ axis in the 
$G$-AF phase to lying almost within the $ab$ planes in the $C$-AF phase, 
with some small alternating $G$-AF-like component.\cite{Ren00}
In addition, the magnetization is strongly reduced at $T>T_{N2}$, being 
only close to $1.0\mu_B$ in the $C$-AF phase,\cite{Kaw94} and the 
magnetic exchange constants $J_{ab}$ and $|J_c|$ are there much lower 
than those found in the low-temperature $G$-AF phase.\cite{Ulr03} 
Even more surprising is the observed gap in the spin wave spectrum which 
suggests an exotic dimerized structure with alternating stronger and 
weaker FM exchange constants along $c$ axis.\cite{Ulr03,Hor03} 
In addition, recent Raman experiments\cite{Miy06} suggest that the 
short-range orbital fluctuations of the orbital $G$-type occur in this 
intermediate $C$-AF phase in addition to the alternating orbital (AO) 
$C$-type ($C$-AO) order, and make it thus quite different from the one 
observed in LaVO$_3$. We also note that the competition between $C$-AF 
and $G$-AF phase is a common feature of a few vanadate compounds with 
low atomic radii.\cite{Miy06}

%%%%%%%%%%%%%%%%%%%%%%%%%%%%%%%%%%%%%%%%%%%%%%%%%%%%%%%%%%%%%%%%%%%%%%%%%
%%                  cubic vanadates: band structure
%%%%%%%%%%%%%%%%%%%%%%%%%%%%%%%%%%%%%%%%%%%%%%%%%%%%%%%%%%%%%%%%%%%%%%%%%
The electronic structure calculations gave valuable information 
about the possible charge distribution over the $t_{2g}$ orbitals in
YVO$_3$.\cite{Saw96,Sol06} Large on-site Coulomb interaction $U$ 
prevents double occuppancy of $d$ orbitals --- it is implemented in the 
calculations using the local density approximation (LDA) within
the so-called LDA+$U$ method.\cite{Ani91} The commonly accepted picture 
is that the $xy$ orbitals are occupied by one electron, while the second 
one occupies either $yz$ or $xz$ orbital.
The lattice distortions in YVO$_3$ are larger in the low-temperature 
phase and suggest $C$-AO order. Above $T_{N2}$ the distortions decrease 
and are compatible with a weak $G$-type AO ($G$-AO) order.\cite{Bla01} 
Theoretical analysis within the charge-transfer model has shown that 
both phases are indeed energetically close,\cite{Miz99} and one may thus 
expect that small changes of the thermodynamic potential around $T_{N2}$ 
could induce a first order phase transition. 

%%%%%%%%%%%%%%%%%%%%%%%%%%%%%%%%%%%%%%%%%%%%%%%%%%%%%%%%%%%%%%%%%%%%%%%%%
%%                    why we spend so much time on it
%%%%%%%%%%%%%%%%%%%%%%%%%%%%%%%%%%%%%%%%%%%%%%%%%%%%%%%%%%%%%%%%%%%%%%%%%
In this paper we study the magnetic properties of cubic vanadates with
a spin-orbital model derived for vanadates some time ago.\cite{Kha01} 
This model applies to Mott insulators with transition metal ions with 
partly filled $t_{2g}$ orbitals in either $d^2$ or $d^4$ configuration. 
Therefore, this model was recently used to analyze the magnetic structure 
of monolayer ruthenates.\cite{Cuo06} In the context of vanadates we 
have already shown before that the orbital fluctuations play a prominent 
role in this model and amplify the FM coupling along the $c$ axis, 
providing a microscopic explanation of the observed $C$-AF order in 
LaVO$_3$. In fact, FM interactions induced by Hund's exchange 
$\propto J_H$ alone are typically much weaker than the AF ones, 
and would not be sufficient to explain why the FM interactions are 
{\it even stronger\/} than the AF ones in the high temperature 
$C$-AF phase of YVO$_3$.  

Here we will concentrate on the exotic magnetic properties of 
YVO$_3$ and address several open questions motivated by the observed 
magnetic properties, in particular why:
  (i) the spin exchange interactions are so different in $G$-type and 
      $C$-type AF phases of YVO$_3$, 
 (ii) the magnetic transition at $T_{N2}$ takes place, 
(iii) the order parameter $\langle S^z\rangle$ in the $C$-AF is so
      strongly reduced, and finally,
 (iv) the dimerization along the $c$ axis, observed in the $C$-AF phase 
      in the intermediate regime of temperature $T_{N2}<T<T_{N1}$, takes 
      place. 
A careful discussion of these questions in the context of the microscopic 
model will lead us to a scenario for the exotic magnetic properties of 
the intermediate temperature phase of YVO$_3$ consistently explained 
within a {\it dimerized\/} $C$-AF order stable only at finite temperature, 
and characterized by reduced exchange interactions. At the same time, 
we will argue that further theoretical studies are necessary in order 
to explain all the observed properties. 
  
%%%%%%%%%%%%%%%%%%%%%%%%%%%%%%%%%%%%%%%%%%%%%%%%%%%%%%%%%%%%%%%%%%%%%%%%%
%%                              outline
%%%%%%%%%%%%%%%%%%%%%%%%%%%%%%%%%%%%%%%%%%%%%%%%%%%%%%%%%%%%%%%%%%%%%%%%%
The paper is organized as follows. In Sec. \ref{sec:model} we present the 
spin-orbital model for cubic vanadates. It is derived from the degenerate 
Hubbard model (Sec. \ref{sec:hub}) and contains superexchange interactions 
supplemented by orbital interactions induced by the lattice (Sec. 
\ref{sec:sex}). Next we introduce the possible types of classical order in 
Sec. \ref{sec:types}, emphasizing first the tendency towards 
one-dimensional (1D) orbital fluctuations (Sec. \ref{sec:eta0}), and next 
comparing their classical energies (Sec. \ref{sec:claso}). The effective 
exchange interactions in different magnetic phases are evaluated in Sec. 
\ref{sec:allj}. For the magnetic phases stable in different regimes of 
parameters we derive spin (Sec. \ref{sec:sw}) and orbital (Sec. 
\ref{sec:ow}) excitations, which serve next to calculate the quantum 
corrections to the energy and lead to the phase diagram of the model at 
$T=0$ of Sec. \ref{sec:phd}. 

Using the above background information we propose a scenario 
for the magnetic phase transition at $T_{N2}$ in YVO$_3$ in Sec. 
\ref{sec:scenario}. The unique instability of the 1D spin-orbital chain 
(Sec. \ref{sec:dim}) comes here together with the reduction of the 
magnetic exchange constants by orbital fluctuations (Sec. \ref{sec:red}) 
to stabilize the dimerized $C$-AF phase at temperature $T>T_{N2}$, as we 
show by analyzing the spin and orbital entropy contributions to the free 
energy (Sec. \ref{sec:trans}). In Sec. \ref{sec:finale} we summarize the 
results and present general conclusions. The paper includes two 
appendices which present the derivation of the spin-orbital model for 
cubic vanadates (Appendix \ref{sec:derivation}), and the calculation of 
spin and orbital excitations, as well as the average order parameters, 
and intersite (spin and orbital) correlations at finite
temperature in the dimerized $C$-AF phase (Appendix \ref{sec:quantum}).

%%%%%%%%%%%%%%%%%%%%%%%%%%%%%%%%%%%%%%%%%%%%%%%%%%%%%%%%%%%%%%%%%%%%%%%%%
%%                               model
%%%%%%%%%%%%%%%%%%%%%%%%%%%%%%%%%%%%%%%%%%%%%%%%%%%%%%%%%%%%%%%%%%%%%%%%%
\section{ Spin-orbital model for cubic vanadates  }
\label{sec:model}

%%%%%%%%%%%%%%%%%%%%%%%%%%%%%%%%%%%%%%%%%%%%%%%%%%%%%%%%%%%%%%%%%%%%%%%%%
%%                        starting with Hubbard
%%%%%%%%%%%%%%%%%%%%%%%%%%%%%%%%%%%%%%%%%%%%%%%%%%%%%%%%%%%%%%%%%%%%%%%%%
\subsection{ Degenerate Hubbard model for $t_{2g}$ electrons}
\label{sec:hub}

We consider a realistic degenerate Hubbard model for $3d$ electrons of 
V$^{3+}$ ions in cubic vanadates, with partly filled $t_{2g}$ orbitals 
that are energetically favored over $e_g$ orbitals by the octahedral field. 
Thereby, we neglect small lattice distortions and the tilting of VO$_6$ 
octahedra. Therefore, the $e_g$ orbitals do not couple to $t_{2g}$ 
orbitals by the hopping processes and play no role in the magnetic 
properties we address below. In such an (idealized) perovskite structure 
V$^{3+}$ ions occupy the cubic lattice, and the hopping elements 
between active $t_{2g}$ orbitals are the 
same in all three cubic directions. The model Hamiltonian,
\begin{equation}
\label{hband}
{\cal H} = H_{t} + H_{\rm cf} + H_{\rm int},
\end{equation}
includes the kinetic energy $H_t$, the orbital splittings induced by 
the crystal field $H_{\rm cf}$, and the on-site electron-electron 
interactions $H_{\rm int}$. The kinetic energy is described by the 
effective hopping element $t$ between two V$^{3+}$ 
ions which originates from two hopping processes via the $2p_{\pi}$ 
oxygen orbital along each Mn--O--Mn bond. Its value can in principle be 
derived from the charge-transfer model,\cite{Zaa93,Miz96} and one 
expects $t=t_{pd}^2/\Delta\sim 0.2$ eV. A more accurate estimation from 
the theory is not possible at the moment, so we will have to rely on 
experimental information from neutrobn scattering concerning the 
magnetic exchange constants in YVO$_3$. 

The kinetic energy is given by:
\begin{eqnarray}
\label{hkin}
H_{t}&=&-t\sum_{\langle ij\rangle{\parallel}\gamma}
         \sum_{\mu(\gamma),\sigma}
\left(d^{\dagger}_{i\mu\sigma}d^{}_{j\mu\sigma}+
      d^{\dagger}_{j\mu\sigma}d^{}_{i\mu\sigma}\right),
\end{eqnarray}
where $d^{\dagger}_{i\nu\sigma}$ are electron creation operators, and 
the summation runs over the bonds $\langle ij\rangle{\parallel}\gamma$ 
along three cubic axes, $\gamma=a,b,c$. As observed before,
\cite{Kha00,Kha01} only two out of three $t_{2g}$ orbitals, labelled by 
$\mu(\gamma)$, are active along each bond $\langle ij\rangle$ and 
contribute to the kinetic energy (\ref{hkin}), while the third orbital 
lies in the plane perpendicular to the $\gamma$ axis and the hopping via 
the $2p_{\pi}$ oxygen is forbidden by symmetry. 
This motivates a convenient notation used below, 
\begin{equation}
\label{abc}
|a\rangle\equiv |yz\rangle, \hskip .7cm
|b\rangle\equiv |xz\rangle, \hskip .7cm
|c\rangle\equiv |xy\rangle,
\end{equation}
with the inactive orbital along a given cubic direction $\gamma$, 
labelled by its index as $|\gamma\rangle$. 

%%%%%%%%%%%%%%%%%%%%%%%%%%%%%%%%%%%%%%%%%%%%%%%%%%%%%%%%%%%%%%%%%%%%%%%%%
%%                        correct interactions
%%%%%%%%%%%%%%%%%%%%%%%%%%%%%%%%%%%%%%%%%%%%%%%%%%%%%%%%%%%%%%%%%%%%%%%%%
The electron-electron interactions are described by the on-site terms,
\cite{Ole83}
\begin{eqnarray}
\label{hint}
H_{\rm int}&=&U\sum_{i\mu}n_{i\mu  \uparrow}n_{i\mu\downarrow}
  +\Big(U-\frac{5}{2}J_H\Big)\sum_{i,\mu<\nu,\sigma\sigma'}
                      n_{i\mu\sigma}n_{i\nu\sigma'}         \nonumber \\
&-&2J_H\sum_{i,\mu<\nu}{\vec S}_{i\mu}\cdot{\vec S}_{i\nu}           
 +  J_H\sum_{i,\mu\neq\nu}
              d^{\dagger}_{i\mu  \uparrow}d^{\dagger}_{i\mu\downarrow}
              d^{       }_{i\nu\downarrow}d^{       }_{i\nu  \uparrow},
                                                            \nonumber \\
\end{eqnarray}
with $U$ and $J_H$ standing for the intraorbital Coulomb and on-site 
Hund's exchange interaction, respectively, using the notation of 
Kanamori.\cite{Kan60} Each pair of orbitals $\{\mu,\nu\}$ is included 
only once in the respective interaction terms with summations over 
$\mu<\nu$. The Hamiltonian (\ref{hint}) describes rigorously the 
multiplet structure of $d^2$ and $d^3$ ions within the $t_{2g}$ 
subspace\cite{Gri71} and is rotationally invariant in the orbital 
space.\cite{Ole83} 
More precisely, the on-site Coulomb interactions depend on three Racah
parameters $\{A,B,C\}$, and for $t_{2g}$ orbitals one finds, 
\begin{equation}
\label{uj}
  U=A+4B+3C, \hskip .7cm J_H=3B+C,                                                   
\end{equation}
The Coulomb and exchange element, $U$ and $J_H$, can be thus obtained 
using the spectroscopic information about the Racah parameters for  
V$^{2+}$ ions in the excited states: $A=3.54$ eV, $B=0.095$ eV, and 
$C=0.354$ eV, as given by Zaanen and Sawatzky.\cite{Zaa90} With these 
parameters one finds $U=5.0$ eV and $J_H=0.64$ eV.

%%%%%%%%%%%%%%%%%%%%%%%%%%%%%%%%%%%%%%%%%%%%%%%%%%%%%%%%%%%%%%%%%%%%%%%%
%%                           ground state
%%%%%%%%%%%%%%%%%%%%%%%%%%%%%%%%%%%%%%%%%%%%%%%%%%%%%%%%%%%%%%%%%%%%%%%%
The Coulomb element $U$ is therefore sufficiently large compared to 
$t\sim 0.2$ eV (i.e., $U\gg t$) to use the second order perturbation 
theory in which the charge fluctuations 
$d_i^2d_j^2\rightleftharpoons d_i^3d_j^1$ are suppressed, and the 
$d$ electrons are localized in $t_{2g}^2$ configurations of a Mott 
insulator (The interaction parameters for V$^{3+}$ ions have similar 
values to those of V$^{2+}$ ones). We use this picture as a starting 
point for our analysis and assume that two electrons are localized at 
each $V^{3+}$ ion $i$, satisfying a local constraint (at site $i$) 
for the total electron density, 
\begin{equation}
\label{const}
n_i=n_{ia}+n_{ib}+n_{ic}=2,
\end{equation}
where $n_{ia}=\sum_{\sigma}n_{ia\sigma}$, {\it etcetera\/}. Two 
electrons at every site are in the high-spin $^3T_2$ triplet ($S=1$) 
state, stabilized by Hund's exchange $J_H$. As $t\ll J_H$,
the kinetic energy $H_{t}$ can only contribute in virtual processes 
which are responsible for the superexchange interactions derived below 
in Sec. \ref{sec:sex}.

\begin{figure}[t!]
\includegraphics[width=7cm]{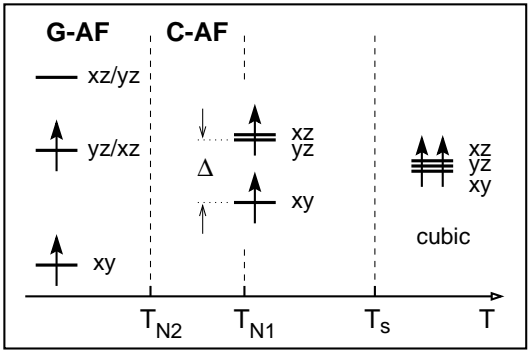}
\caption{
An artist view of the energy splittings between $t_{2g}$ orbitals in 
YVO$_3$ in different temperature regimes. The orbital splitting 
$\Delta$ which occurs below the structural transition at $T_s$ and 
persists in the $C$-AF phase favors the occupied $xy$ orbitals, but 
allows also for weak orbital fluctuations. Such fluctuations are 
quenched in the $G$-AF phase at $T<T_{N2}$.  
} 
\label{fig:artist}
\end{figure}

%%%%%%%%%%%%%%%%%%%%%%%%%%%%%%%%%%%%%%%%%%%%%%%%%%%%%%%%%%%%%%%%%%%%%%%%
%%                           crystal field
%%%%%%%%%%%%%%%%%%%%%%%%%%%%%%%%%%%%%%%%%%%%%%%%%%%%%%%%%%%%%%%%%%%%%%%%
The third term in Eq. (\ref{hband}) stands for the orbital energies in
crystal field induced by the structural transition at $T_s\sim 200 K$,
\cite{Bla01} which lifts the degeneracy of three $t_{2g}$ orbitals and 
breaks the cubic symmetry in the orbital space. We write the crystal 
field term $H_{\rm cf}$ as follows,
\begin{equation}
\label{hpot}
H_{\rm cf}=\sum_{i\mu}\varepsilon_{i\mu}n_{i\mu},
\end{equation}
with electron energies $\varepsilon_{i\mu}$ for orbital $\mu$ at site 
$i$. In agreement with the results of band structure calculations,
\cite{Saw96,Sol06} and with an idealized but suggested by the local 
distortions and thus commonly accepted picture,\cite{Ren00} we assume 
that the $xy$ orbitals are favored below the structural transition, 
while the remaining $yz$ and $xz$ orbitals are nearly degenerate, i.e., 
$\varepsilon_c<\varepsilon_a$, and $\varepsilon_b\simeq\varepsilon_a$,
leading to
\begin{equation}
\label{frozen}
n_{ic}\simeq 1, \hskip .7cm  n_{ia}+n_{ib}\simeq 1,
\end{equation}
i.e., $c$ orbitals are 'condensed' and the other two represent the 
remaining $t_{2g}$ orbital degree of freedom at every site 
(see Fig. \ref{fig:artist}). Although in principle the orbital energies 
$\varepsilon_{i\mu}$ could change at the magnetic transition at $T_{N2}$ 
and further stabilize $G$-AF phase at low temperature, we will ignore 
small corrections which would result from this effect in the derivation
of the superexchange, and consider only generic features of the 
spin-orbital model that could be responsible for the experimental 
situation. 

%%%%%%%%%%%%%%%%%%%%%%%%%%%%%%%%%%%%%%%%%%%%%%%%%%%%%%%%%%%%%%%%%%%%%%%%
%%                             Figure 1
%%%%%%%%%%%%%%%%%%%%%%%%%%%%%%%%%%%%%%%%%%%%%%%%%%%%%%%%%%%%%%%%%%%%%%%%

%%%%%%%%%%%%%%%%%%%%%%%%%%%%%%%%%%%%%%%%%%%%%%%%%%%%%%%%%%%%%%%%%%%%%%%%
%%                       Superexchange interactions
%%%%%%%%%%%%%%%%%%%%%%%%%%%%%%%%%%%%%%%%%%%%%%%%%%%%%%%%%%%%%%%%%%%%%%%%
\subsection{ Superexchange model for vanadates }
\label{sec:sex}

Consider first the atomic limit, i.e., the system of $V^{3+}$ ions in 
$d^2$ configuration at $t=0$. In the ground state $S=1$ spin forms at 
each ion, and one finds a large degeneracy $9^N$ of the ground state, 
where $N$ is the number of sites, as every spin component 
($S^z=1,0,-1$) is allowed, and a hole may occupy either orbital: 
$|a\rangle$, $|b\rangle$ or $|c\rangle$. This large degeneracy is, 
however, removed by the effective interactions between each pair of 
nearest neighbor ions $\{i,j\}$, which originate from virtual 
transitions to the excited states due to charge  
$d^2_id^2_j\rightleftharpoons d^3_id^1_j$ excitations, generated in each 
case by a single hopping of a $t_{2g}$ electron. In the realistic regime 
of parameters such processes may be treated perturbatively, and one 
arrives in second order perturbation theory at an effective 
superexchange Hamiltonian of Ref. \onlinecite{Kha01} --- the details of 
the derivation are explained in Appendix \ref{sec:derivation}. 

\begin{figure}[t!]
\includegraphics[width=8cm]{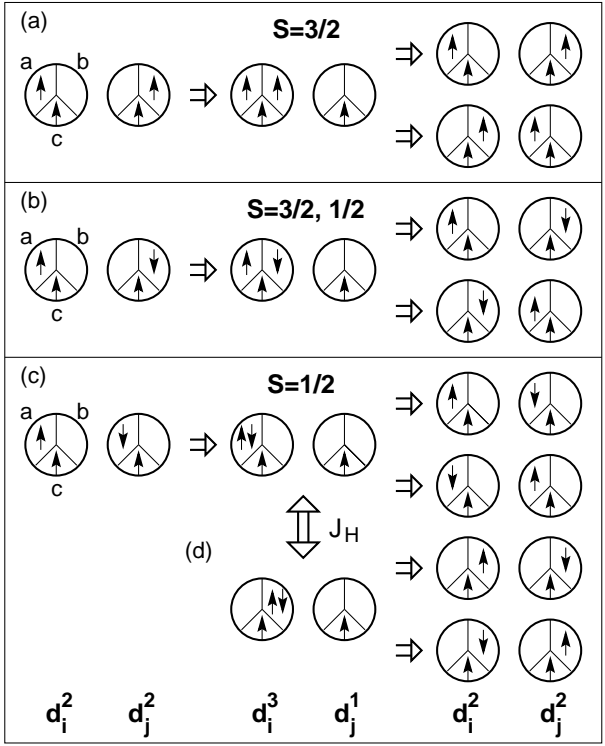}
\caption{
Virtual charge excitations $d^2_id^2_j\rightarrow d^3_id^1_j\rightarrow 
d^2_id^2_j$ within a bond $\langle ij\rangle$ along $c$ axis, which 
contribute to the superexchange in cubic vanadates. Orbital
fluctuations which support the FM superexchange occur when different 
active orbitals $a$ and $b$ are occupied at both sites, as in cases 
(a) and (b). If the same orbitals are occupied at both sites, e.g. the 
orbital $a$ as in case (c), the superexchange is AF  --- then 
a double occupancy of the occupied (and active) orbital is generated in 
the excited state, which next dissociates to a configuration with either: 
(c) the same orbital occupancies, or 
(d) with interchanged occupied orbitals at sites $i$ and $j$. 
} 
\label{fig:se}
\end{figure}

%%%%%%%%%%%%%%%%%%%%%%%%%%%%%%%%%%%%%%%%%%%%%%%%%%%%%%%%%%%%%%%%%%%%%%%%
%%                        our spin-orbital model 
%%%%%%%%%%%%%%%%%%%%%%%%%%%%%%%%%%%%%%%%%%%%%%%%%%%%%%%%%%%%%%%%%%%%%%%%
The superexchange interactions between two $S=1$ spins at sites $i$ and 
$j$ arise from virtual excitations $d^2_id^2_j\rightarrow d^3_id^1_j$ 
along the concerned bond $\langle ij\rangle$, promoted by the hopping 
$t$ which couples pairs of identical {\it active $t_{2g}$ orbitals\/}. 
A single hopping process generates a $d^3_i$ configuration, either with 
three different orbitals occupied by a single electron each, or with a 
double occupancy in one of the two active orbitals (see Fig. \ref{fig:se}). 
Therefore, the $d^3_i$ excited state may be either a high-spin $^4A_2$ 
state, or one of three low-spin states: $^2E$, $^2T_1$ or $^2T_2$ with 
energies\cite{notex} $U-3J_H$, $U$ and $U+2J_H$, as shown in Fig. 1 of 
Ref. \onlinecite{Ole05}. This perturbative consideration leads to the 
spin-orbital superexchange model for cubic vanadates,   
\begin{equation}
\label{HJ}
{\cal H}_J=J\sum_{\langle ij\rangle\parallel\gamma}\left[ 
\left({\vec S}_i\cdot {\vec S}_j+1\right)
      {\hat J}_{ij}^{(\gamma)} + 
      {\hat K}_{ij}^{(\gamma)}    \right].
\end{equation}
with the energy scale given by the superexchange constant, 
\begin{equation}
\label{sex}
J=\frac{4t^2}{U}.
\end{equation}
The spin interactions $\propto {\vec S}_i\cdot {\vec S}_j$ obey the 
SU(2) symmetry. In contrast, the orbital interaction operators
${\hat J}_{ij}^{(\gamma)}$ and ${\hat K}_{ij}^{(\gamma)}$ involve only 
two active $t_{2g}$ orbitals on each individual bond 
$\langle ij\rangle\parallel\gamma$ ($\gamma=a,b,c$) which contribute 
to the virtual excitations, so they have a lower (cubic) symmetry. 
These operators take the form:
\begin{eqnarray}
\label{orbj}
{\hat J}_{ij}^{(\gamma)}&=&\!
\frac{1}{2}\left[(1+2\eta r_1)
\left({\vec\tau}_i\cdot {\vec\tau}_j
     +\frac{1}{4}n_i^{}n_j^{}\right)\right.          \nonumber \\ 
&-&\! \left.\eta r_3
    \left({\vec\tau}_i\otimes{\vec\tau}_j+\frac{1}{4}n_i^{}n_j^{}\right)
-\frac{1}{2}\eta r_1(n_i+n_j)\right]^{(\gamma)}\!,             \\ 
\label{orbk}
{\hat K}_{ij}^{(\gamma)}&=&\!
\left[\eta r_1
\left({\vec\tau}_i\cdot {\vec\tau}_j+\frac{1}{4}n_i^{}n_j^{}\right)  
 +\eta r_3\left({\vec\tau}_i\otimes {\vec\tau}_j
             +\frac{1}{4}n_i^{}n_j^{}\right)\right.  \nonumber \\ 
&-&\!\left. 
   \frac{1}{4}(1+\eta r_1)(n_i+n_j)\right]^{(\gamma)}\!,
\end{eqnarray}
and have a rich structure which originates from the projections of the 
$d^3_i$ excited states on the respective eigenstates of V$^{2+}$ ion, 
as explained in Appendix \ref{sec:derivation}.  

%%%%%%%%%%%%%%%%%%%%%%%%%%%%%%%%%%%%%%%%%%%%%%%%%%%%%%%%%%%%%%%%%%%%%%%%
%%                              Figure 2
%%%%%%%%%%%%%%%%%%%%%%%%%%%%%%%%%%%%%%%%%%%%%%%%%%%%%%%%%%%%%%%%%%%%%%%%

%%%%%%%%%%%%%%%%%%%%%%%%%%%%%%%%%%%%%%%%%%%%%%%%%%%%%%%%%%%%%%%%%%%%%%%%
%%                            notation
%%%%%%%%%%%%%%%%%%%%%%%%%%%%%%%%%%%%%%%%%%%%%%%%%%%%%%%%%%%%%%%%%%%%%%%%
First of all, the interactions ${\hat J}_{ij}^{(\gamma)}$ and 
${\hat K}_{ij}^{(\gamma)}$ depend on Hund's exchange splittings in the 
multiplet structure of a V$^{2+}$ ion in local $d^3$ configuration 
(shown in Fig. 1 of Ref. \onlinecite{Ole05}) via the exchange parameter,
\begin{equation}
\label{eta}
\eta=\frac{J_H}{U}, 
\end{equation}
and the respective coefficients $r_1$ and $r_3$  in Eqs. (\ref{orbj}) and 
(\ref{orbk}) are (for convenience we use here the same notation as in Ref. 
\onlinecite{Ole05}): 
\begin{equation}
\label{rr}
r_1=\frac{1}{1-3\eta}, \hskip 1.5cm r_3=\frac{1}{1+2\eta}.
\end{equation}
They correspond to the excitation spectrum in     
$d_i^2d_j^2\rightleftharpoons d_i^3d_j^1$ charge transitions (Fig. \ref{fig:se}). 
In the present case of cubic vanadates one finds\cite{Zaa90} $\eta\simeq 0.13$, 
which we will take as a representative value for YVO$_3$. 

The pseudospin (orbital) operators ${\vec\tau}_i=\{\tau_i^+,\tau_i^-,
\tau_i^z\}$ for pseudospin $\tau=1/2$ in Eqs. (\ref{orbj}) and 
(\ref{orbk}) are defined in the subspace spanned by two orbital flavors 
which are active along a given direction $\gamma$. 
For instance, the virtual transitions which generate the superexchange 
interactions follow from the electron hopping between the pairs of 
active $a$ and $b$ orbitals along the bond $\langle ij\rangle\parallel c$ 
axis (see Fig. \ref{fig:se}), and these operators are defined by Eqs. 
(\ref{sextau}), while the number of active electrons at site $i$ is 
$n_i^{(c)}=n_{ia}^{}+n_{ib}^{}$. It is important to realize that 
although the pseudospin flavor is conserved in each individual hopping 
processes, the off-diagonal elements of the Coulomb interaction 
$\propto J_H$ generate transitions between the components of the excited 
states, as shown in Fig. \ref{fig:se}(d).\cite{noteo} Therefore, next to 
the usual scalar products,
\begin{equation}
\label{heis}
2\left({\vec\tau}_i\cdot {\vec\tau}_j\!
+\frac{1}{4}n_i^{}n_j^{}\right)^{(c)}\equiv (n_{ia}^{}n_{ja}^{}
+a_i^{\dagger}b_i^{}b_j^{\dagger}a_j^{})+(a \leftrightarrow b), 
\end{equation}
we also find in the orbital operators ${\hat J}_{ij}^{(\gamma)}$ and 
${\hat K}_{ij}^{(\gamma)}$ 'orbital fluctuating' terms
\begin{equation}
\label{exo}
2\left({\vec\tau}_i\otimes {\vec\tau}_j\!
      +\frac{1}{4}n_i^{}n_j^{}\right)^{(c)}\equiv 
(n_{ia}^{}n_{ja}^{}+a_i^{\dagger}b_i^{}a_j^{\dagger}b_j^{})+
(a \leftrightarrow b), 
\end{equation}
where $(a \leftrightarrow b)$ stands for the terms with interchanged 
$a$ and $b$ orbitals. Unlike in the Heisenberg model, the interactions 
$\propto\tau_i^+\tau_j^+=a_i^{\dagger}b_i^{}a_j^{\dagger}b_j^{}$ in Eq. 
(\ref{exo}) induce {\it similar\/} orbital flips at both sites. 
Such terms have the form 
\begin{equation}
\label{exo1}
\left({\vec\tau}_i\otimes {\vec\tau}_j\right)^{(c)}=
\frac{1}{2}\left(\tau_i^+\tau_j^++\tau_i^-\tau_j^-\right)
+\tau_i^z\tau_j^z 
\end{equation}
and lead to the nonconservation of the total pseudospin quantum number 
and are thus responsible for further enhancement of orbital quantum 
fluctuations on the bonds with both orbitals active (in this case along 
$c$ axis). In contrast, the bonds in $ab$ planes are classical as there
analogous terms cannot contribute when the $c$ orbitals have condensed. 
This demonstrates that the breaking of symmetry in the orbital space,
such as given by Eqs. (\ref{frozen}), will have severe consequences for 
magnetism.  

The complete microscopic model we consider in the following Sections,
\begin{equation}
\label{model}
{\cal H}={\cal H}_J+{\cal H}_{\rm orb},
\end{equation}
includes as well effective orbital interactions induced by the oxygen 
distortions. When the VO$_6$ octahedra distort at a second magnetic 
transition at $T_{N2}$,\cite{Kaw94,Ree06} intersite interactions which 
help to order $yz$ and $xz$ orbitals, occupied by one electron et every 
site, are induced. They are of two types ---
the GdFeO$_3$-type distortions favor repeated orbitals along the $c$ 
axis, while the AO order in $ab$ planes is favored by weak JT effect. 
Therefore, in addition to the superexchange (\ref{HJ}) we introduce 
two effective orbital interactions $\{V_c,V_a\}$ as the last term of 
the effective Hamiltonian (\ref{model}),
\begin{equation}
\label{Horb}
{\cal H}_{\rm orb}=
-V_c\sum_{\langle ij\rangle\parallel c} \tau_i^z\tau_j^z
+V_a\sum_{\langle ij\rangle\parallel ab}\tau_i^z\tau_j^z,
\end{equation}
where the orbital pseudospin operator $\tau_i^z$ at site $i$ is defined 
by Eq. (\ref{tauz}). With the present sign convention both parameters 
are positive ($V_c>0$ and $V_a>0$) and induce the $C$-type AO ($C$-AO) 
order, as observed in the $G$-AF phase at $T<T_{N_2}$. For convenience 
we express the orbital interactions in ${\cal H}_{\rm orb}$ (\ref{Horb}) 
in the units of the superexchange constant $J$, and introduce 
dimensionless parameters:
\begin{equation}
\label{vac}
v_a=\frac{V_a}{J}\hskip .7cm v_c=\frac{V_c}{J},
\end{equation}
which describe the model given by Eq. (\ref{model}), in addition to 
Hund's exchange parameter $\eta$.

%%%%%%%%%%%%%%%%%%%%%%%%%%%%%%%%%%%%%%%%%%%%%%%%%%%%%%%%%%%%%%%%%%%%%%%% 
%%                        Types of magnetic order
%%%%%%%%%%%%%%%%%%%%%%%%%%%%%%%%%%%%%%%%%%%%%%%%%%%%%%%%%%%%%%%%%%%%%%%% 
\section{ Types of magnetic order }
\label{sec:types}

%%%%%%%%%%%%%%%%%%%%%%%%%%%%%%%%%%%%%%%%%%%%%%%%%%%%%%%%%%%%%%%%%%%%%%%% 
%%                          limit of J_H -> 0 
%%%%%%%%%%%%%%%%%%%%%%%%%%%%%%%%%%%%%%%%%%%%%%%%%%%%%%%%%%%%%%%%%%%%%%%% 
\subsection{ Orbital singlets at $J_H\to 0$ }
\label{sec:eta0}

In order to understand the possible symmetry breaking in the cubic 
vanadates, consider first the superexchange interactions in the 
$J_H\to 0$ limit:
\begin{equation}
{\cal H}_0=\frac{1}{2}J\sum_{\langle ij\rangle\parallel\gamma}
\left({\vec S}_i\cdot {\vec S}_j+1\right)
\left({\vec\tau}_i\cdot {\vec\tau}_j
+\frac{1}{4}n_i^{}n_j^{}\right)^{(\gamma)},
\label{pauli}
\end{equation}
where a constant energy of $-2J$ per V$^{3+}$ ion is neglected. It is 
straighforward to understand why the interactions at $J_H\to 0$ turn out 
to have the same structure as in LaTiO$_3$,\cite{Kha00} where for spins 
$s=1/2$ of Ti$^{3+}$ ions one finds instead the spin part 
$4({\vec s}_i\cdot {\vec s}_j+\frac{1}{4})$. In fact, in the limit of 
$J_H\to 0$ the superexchange interactions follow entirely from the 
{\it Pauli principle\/}, as the multiplet structure of excited states 
collapses to a single degenerate level and the spin interactions
$\propto{\vec S}_i\cdot{\vec S}_j$ due to the high-spin $^4A_2$ and 
low-spin $^2E$ states, which involve $d^3\{abc\}$ configurations, cancel 
each other (see Appendix \ref{sec:derivation}). This suggests that the 
superexchange interactions might {\it all\/} be AF in the limit of 
$J_H\to 0$, as in $e_g$ systems.\cite{Fei97,Ole01} In fact, in $e_g$ 
systems only one directional orbital is active along the bond, two 
electrons occupying these orbitals form an intraorbital {\it spin singlet\/}, 
which maximizes the energy gain for the AF superexchange. 

\begin{figure}[t!]
\includegraphics[width=8cm]{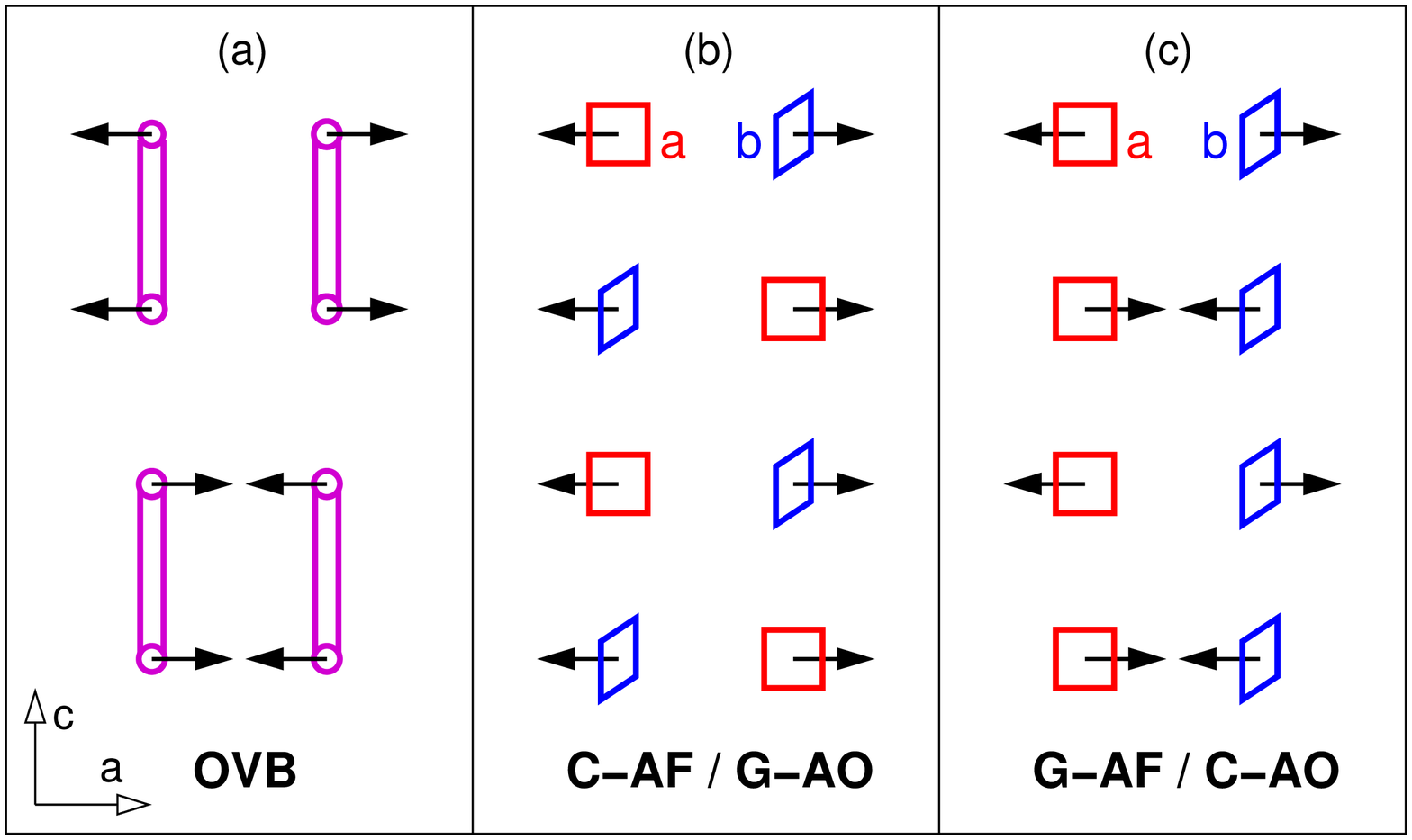}
\caption{(Color online)
Schematic picture of the classical phases with magnetic and orbital
order for $n_{ic}=1$ in $ac$ plane:  
(a) OVB phase, with alternating strong FM bonds stabilized by orbital 
    singlets represented by double lines, and weak AF bonds (with 
    disordered $a/b$ orbitals);
(b) $C$-AF spin order accompanied by $G$-AO order; and  
(c) $G$-AF spin order accompanied by $C$-AO order with repeated either 
    $a$ or $b$ orbitals along $c$ axis. 
In cases (b) and (c) spins and orbitals alternate along $b$ direction
(not shown). These latter states follow the Goodenough-Kanamori rules
\cite{Goo63,Kan59} and are analyzed below for YVO$_3$.
} 
\label{fig:claso}
\end{figure}

%%%%%%%%%%%%%%%%%%%%%%%%%%%%%%%%%%%%%%%%%%%%%%%%%%%%%%%%%%%%%%%%%%%%%%%%%
%%                          orbital singlets
%%%%%%%%%%%%%%%%%%%%%%%%%%%%%%%%%%%%%%%%%%%%%%%%%%%%%%%%%%%%%%%%%%%%%%%%%
However, there is an important difference between the $e_g$ (with one 
hole per site) and $t_{2g}$ (with one or two electrons per site) 
systems, which may be best realized by considering a single bond 
$\langle ij\rangle$ in one cubic direction. Two active $t_{2g}$ orbitals 
along this bond open a new possibility --- if both orbitals are singly 
occupied, an {\it orbital singlet\/} gives here FM superexchange, even in 
the absence of Hund's exchange $J_H$.\cite{Kha01} For the present filling 
of $n=2$ electrons per site, and if $n_c=1$, such a resonance on a bond 
is possible only along one out of three cubic directions\cite{notet} 
--- the orbital singlets and uncorrelated bonds alternate along the $c$ 
axis [Fig. \ref{fig:claso}(a)]. In analogy with spin systems,\cite{Koh88} 
this state can be called an {\it orbital valence bond\/} (OVB) state.
\cite{Ole02} This possibility was also independently pointed out by Shen, 
Xie and Zhang,\cite{She02} who obtained the OVB state as the most stable 
solution of the present Hamiltonian (\ref{HJ}) in the regime of small 
$\eta$ and for large $S$ limit. 

%%%%%%%%%%%%%%%%%%%%%%%%%%%%%%%%%%%%%%%%%%%%%%%%%%%%%%%%%%%%%%%%%%%%%%%%%
%%                        magnetism at eta=0
%%%%%%%%%%%%%%%%%%%%%%%%%%%%%%%%%%%%%%%%%%%%%%%%%%%%%%%%%%%%%%%%%%%%%%%%%
The OVB state implies an unconventional type of magnetic order. 
At $\eta=0$ the exchange constants along the $c$ axis are given by 
\begin{equation}
\label{J0}
J_{c}(\eta=0)=\left\langle{\vec\tau}_i\cdot {\vec\tau}_j
+\frac{1}{4}n_i^{}n_j^{}\right\rangle^{(c)}.
\end{equation}
When the orbital singlets form and contribute to the energy with 
$\langle{\vec\tau}_i\cdot {\vec\tau}_j\rangle^{(c)}=-\frac{3}{4}$, they
maximize the FM exchange on these bonds [see Sec. \ref{sec:allj}], 
and stabilize there effective $S=2$ spin states.
Between them one finds disordered orbitals, i.e., 
$\langle{\vec\tau}_i\cdot{\vec\tau}_j\rangle^{(c)}=0$, 
so the magnetic exchange interactions on these bonds are much weaker and 
are in fact AF due to the static term $\langle n_in_j\rangle^{(c)}=1$ in 
Eq. (\ref{J0}). The interactions within the $ab$ planes are also AF but 
somewhat stronger --- they follow from the conventional (Pauli principle) 
mechanism which operates as well in the absence of orbital degeneracy, 
with {\it intraorbital\/} singlets generated by the nearest neighbor 
hopping between sites with singly occupied $c$ orbitals (\ref{frozen}). 
Assuming disordered orbitals one finds 
$\langle{\vec\tau}_i\cdot {\vec\tau}_j\rangle^{(ab)}=0$ and 
$\langle n_in_j\rangle^{(ab)}=\frac{5}{2}$ for the bonds in $ab$ planes. 
The resulting magnetic order which coexists with the 
orbital singlets is shown in Fig. \ref{fig:claso}(a).

%%%%%%%%%%%%%%%%%%%%%%%%%%%%%%%%%%%%%%%%%%%%%%%%%%%%%%%%%%%%%%%%%%%%%%%%
%%                             Figure 3
%%%%%%%%%%%%%%%%%%%%%%%%%%%%%%%%%%%%%%%%%%%%%%%%%%%%%%%%%%%%%%%%%%%%%%%%

%%%%%%%%%%%%%%%%%%%%%%%%%%%%%%%%%%%%%%%%%%%%%%%%%%%%%%%%%%%%%%%%%%%%%%%%%
%%                  Phases with magnetic and orbital order       
%%%%%%%%%%%%%%%%%%%%%%%%%%%%%%%%%%%%%%%%%%%%%%%%%%%%%%%%%%%%%%%%%%%%%%%%%
\subsection{ Magnetic and orbital order at finite $J_H$ }
\label{sec:claso}

Let us analyze the possible types of coexisting magnetic and orbital 
order of the full effective Hamiltonian given by Eq. (\ref{model}) for 
finite Hund's exchange $J_H$, which includes the effective orbital 
interactions $\{V_a,V_c\}$ with the lattice. Motivated by the 
experimental situation in YVO$_3$, we assume that the $c$ orbitals have 
condensed, so the constraints given by Eqs. (\ref{frozen}) are fulfilled. 
For this case we consider possible classical phases and their energies. 
A more complete analysis which includes the quantum corrections due to 
spin and orbital excitations is presented in Sec. \ref{sec:phd}, here we 
discuss only a qualitative picture when spin quantum fluctuations are 
neglected.

%%%%%%%%%%%%%%%%%%%%%%%%%%%%%%%%%%%%%%%%%%%%%%%%%%%%%%%%%%%%%%%%%%%%%%%%%
%%                             OVB order
%%%%%%%%%%%%%%%%%%%%%%%%%%%%%%%%%%%%%%%%%%%%%%%%%%%%%%%%%%%%%%%%%%%%%%%%%
At $\eta=0$ the lowest energy is obtained when the orbital fluctuations 
are fully developed at every second bond along the $c$ axis in the OVB
state,\cite{Ulr03,She02,Hor03} as discussed in Sec. \ref{sec:eta0}. The 
classical energy of this phase per site is 
obtained assuming the classical values for intersite spin correlations: 
$\langle {\vec S}_i\cdot{\vec S}_j\rangle=\pm 1$ on the FM/AF bonds. 
It includes the orbital fluctuation energy gained on the orbital 
singlet bonds and is given by: 
\begin{equation}
\label{eovb}
E_{\rm OVB}^{(0)}=
-J\left[\frac{1}{4}r_1+\frac{1}{8}\eta(9r_1-11r_3)-\frac{1}{8}v_c\right].
\end{equation}
Here and below we neglect a constant nonmagnetic term $-2J$. 
Except for the orbital singlets, the orbital interactions in 
${\cal H}_{\rm orb}$ do not contribute as the $\{a,b\}$ orbitals are 
disordered on all other bonds [Fig. \ref{fig:claso}(a)].

%%%%%%%%%%%%%%%%%%%%%%%%%%%%%%%%%%%%%%%%%%%%%%%%%%%%%%%%%%%%%%%%%%%%%%%%
%%                            C-AF order
%%%%%%%%%%%%%%%%%%%%%%%%%%%%%%%%%%%%%%%%%%%%%%%%%%%%%%%%%%%%%%%%%%%%%%%%
An alternative AF state, realized at larger values of $\eta$,
\cite{Kha01} is obtained when the (negative) orbital correlations along 
the $c$ axis are uniform, and all the bonds exhibit FM exchange. As the 
spin interactions remain AF in $ab$ planes, these interactions lead to 
the $C$-AF phase shown schematically in Fig. \ref{fig:claso}(b). 
A straightforward estimate of the classical energy of this phase,  
\begin{eqnarray}
\label{ecaf}
E_{C}^{(0)}&=&J\Big[
r_1\Big\langle{\vec\tau}_i\cdot{\vec\tau}_j+\frac{1}{4}\Big\rangle^{(c)}
    +\eta(r_1+r_3)\langle n_{ia}n_{ja}\rangle^{(b)}        \nonumber \\
  &-&\eta(2r_1-r_3)-v_c+2v_a\Big],
\end{eqnarray}
is obtained again taking the classical spin correlations: 
$\langle {\vec S}_i\cdot{\vec S}_j\rangle^{(c)}=1$ and 
$\langle {\vec S}_i\cdot{\vec S}_j\rangle^{(ab)}=-1$. It depends on the 
orbital correlations $\langle{\vec\tau}_i\cdot{\vec\tau}_j\rangle$. 
Taking fully disordered 1D orbital chain with 
$\langle{\vec\tau}_i\cdot{\vec\tau}_j\rangle=-0.4431$, as for the AF 
Heisenberg spin chain,\cite{Mat81} one finds a crossover from the OVB 
to the $C$-AF phase at $\eta_0\simeq 0.064$. We improve this naive 
estimate of the transition in Sec. \ref{sec:phd}, where we evaluate the 
quantum corrections due to spin excitations in both phases. 

%%%%%%%%%%%%%%%%%%%%%%%%%%%%%%%%%%%%%%%%%%%%%%%%%%%%%%%%%%%%%%%%%%%%%%%%
%%                            G-AF order
%%%%%%%%%%%%%%%%%%%%%%%%%%%%%%%%%%%%%%%%%%%%%%%%%%%%%%%%%%%%%%%%%%%%%%%%
Unlike in $e_g$ systems,\cite{Fei99,Oka02} the orbital interactions 
induced by the lattice (\ref{Horb}) compete with the superexchange 
(\ref{HJ}) in the present effective spin-orbital model (\ref{model}) 
and stabilize the $G$-AF phase at sufficiently large orbital 
interaction $V_c$.\cite{Kha01} The classical energy of this phase,  
\begin{equation}
\label{egaf}
E_{G}^{(0)} =-J\left[\eta(r_1-r_3)+\frac{1}{4}v_c+\frac{1}{2}v_a\right],
\end{equation}
is lowered by the energy $-\frac{1}{4}J(v_c+2v_a)$ gained per site when 
the $C$-AO order shown in Fig. \ref{fig:claso}(c) sets in. In fact, 
the $C$-AO order enforces here the $G$-AF phase, showing a close 
interrelation of spin and orbital intersite correlations, known in
the literature as the Goodenough-Kanamori rules.\cite{Goo63,Kan59}

%%%%%%%%%%%%%%%%%%%%%%%%%%%%%%%%%%%%%%%%%%%%%%%%%%%%%%%%%%%%%%%%%%%%%%%%
%%                  Excitations and quantum fluctuations     
%%%%%%%%%%%%%%%%%%%%%%%%%%%%%%%%%%%%%%%%%%%%%%%%%%%%%%%%%%%%%%%%%%%%%%%%
\section{ Spin and orbital excitations }
\label{sec:allex}

%%%%%%%%%%%%%%%%%%%%%%%%%%%%%%%%%%%%%%%%%%%%%%%%%%%%%%%%%%%%%%%%%%%%%%%%
%%                            G-AF order
%%%%%%%%%%%%%%%%%%%%%%%%%%%%%%%%%%%%%%%%%%%%%%%%%%%%%%%%%%%%%%%%%%%%%%%%
\subsection{ Effective exchange interations }
\label{sec:allj}

In order to analyze the spin and orbital excitations, we follow the 
usual approach in mean field theory\cite{Ole05} and decouple spin 
and orbital operators in Eq. (\ref{HJ}). Note that this approach is 
satisfactory below the spin ordering temperature $T_{N1}$, as then 
the spin fluctuations are quenched and the spin and orbital degrees 
of freedom may be disentangled,\cite{Ole06} while for $T>T_{N1}$ 
composite spin-orbital excitations need to be considered. This 
procedure leads to the effective spin exchange constants $J_c$ and 
$J_{ab}$, as given in Refs. \onlinecite{Kha04,Ole05},
\begin{eqnarray}
\label{jcd2}
J_c&=&
-\frac{1}{2}J\Big[\eta r_1-(r_1-\eta r_1-\eta r_3)       \nonumber \\
&&\times\Big\langle\vec\tau_i\cdot\vec\tau_j+\frac{1}{4}\Big\rangle^{(c)}
 -2\eta r_3\Big\langle\tau_i^y\tau_j^y\Big\rangle^{(c)}\Big],      \\
\label{jabd2}
J_{ab}&=&\frac{1}{4}J\Big[1-\eta r_1-\eta r_3            \nonumber \\
&&+(r_1-\eta r_1-\eta r_3)
\Big\langle\tau_i^z\tau_j^z+\frac{1}{4}\Big\rangle^{(a)}\Big].
\end{eqnarray}
They depend on the orbital correlations, 
$\langle\vec\tau_i\!\cdot\!\vec\tau_j\rangle$ and 
$\langle\tau_i^y\tau_j^y\rangle$ along $c$ axis, and 
$\langle\tau_i^z\tau_j^z\rangle$ in $ab$ planes, which have to be 
determined from the full superexchange model given by Eq. 
(\ref{model}), i.e., in presence of orbital interactions promoted 
by the lattice. Below we specify the effective exchange interactions 
for three possible phases shown in Fig. \ref{fig:claso}.

\begin{figure}[t!]
\includegraphics[width=7cm]{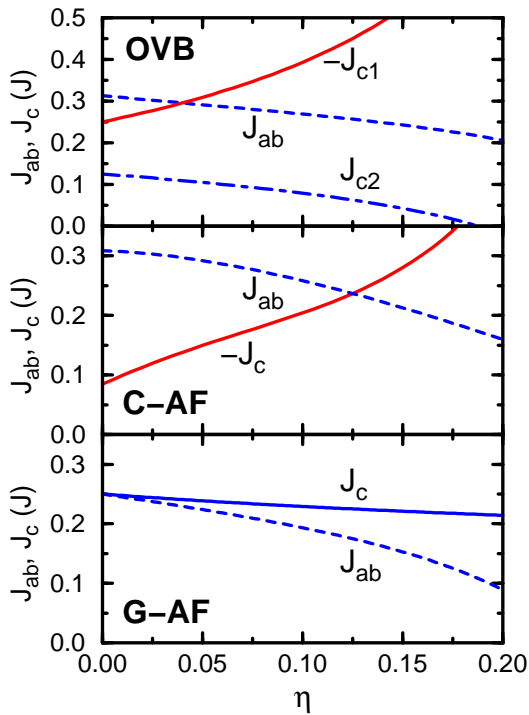}
\caption{(Color online)
Exchange interactions $J_c$ and $J_{ab}$ 
as functions of Hund's exchange $\eta$ as obtained for:  
(a) OVB phase with alternating strong FM $J_{c1}$ (solid line) and weak 
    AF $J_{c2}$ (dashed-dotted line) exchange interaction;
(b) $C$-AF phase with (weak) $G$-type OO; and  
(c) $G$-AF phase stabilized by the orbital interactions $\{V_a,V_c\}$ 
    which induce the $C$-AO order.
FM (AF) exchange interactions along $c$ axis in $C$-AF ($G$-AF) phase 
are shown by solid lines, while 
AF interactions in $ab$ planes are shown by dashed lines.      
} 
\label{fig:ex}
\end{figure}

At low $\eta$ one expects that the OVB state with alternating FM and 
AF bonds along $c$ axis is stable [Fig. \ref{fig:claso}(a)]. On the 
bonds occupied by orbital singlets, with 
$\langle\vec\tau_i\cdot\vec\tau_j\rangle^{(c)}=-\frac{3}{4}$ and 
$\langle\tau_i^y\tau_j^y\rangle^{(c)}=-\frac{1}{4}$, one finds 
strong FM exchange
\begin{equation}
\label{djc1}
J_{c1}^O = -\frac{1}{4}Jr_1(1+\eta),
\end{equation}
which is further enhanced with increasing $\eta$ and soon becomes the
dominating magnetic interaction, see Fig. \ref{fig:ex}(a). In contrast, 
for the bonds connecting singlets the orbitals are disordered,
$\langle\vec\tau_i\cdot\vec\tau_j\rangle=
\langle\tau_i^y\tau_j^y\rangle=0$, and the resulting AF exchange 
interactions
\begin{equation}
\label{djc2}
J_{c2}^O = \frac{1}{8}J[1-\eta(2r_1+r_3)],
\end{equation}
decrease with increasing $\eta$. These exchange interactions are much 
weaker than the AF ones in the $ab$ planes, 
\begin{equation}
\label{djab}
J_{ab}^O = \frac{5}{16}J[1-\eta(r_1+r_3)],
\end{equation}
in the entire allowed regime of $\eta$, as the latter interactions are 
supported by the excitations of doubly occupied configurations in $c$ 
orbitals. One finds that the OVB state with alternating FM and AF bonds 
is destroyed at a critical value of $\eta$,
\begin{equation}
\label{criteta}
\eta_0 =\frac{1}{2r_1+r_3}\simeq 0.188,
\end{equation}
where the weaker AF bond $J_{c2}$ collapses and changes its sign, see 
Fig. \ref{fig:ex}(a). In reality, it turns out that the orbital singlets 
are destabilized even much faster as a better energy is obtained when the 
spins reorient to FM order and the $C$-AF phase with uniform disordered 
(or weakly ordered) $\{a,b\}$ orbitals along the $c$ axis takes over, 
as we show below. 

%%%%%%%%%%%%%%%%%%%%%%%%%%%%%%%%%%%%%%%%%%%%%%%%%%%%%%%%%%%%%%%%%%%%%%%%
%%                             Figure 4
%%%%%%%%%%%%%%%%%%%%%%%%%%%%%%%%%%%%%%%%%%%%%%%%%%%%%%%%%%%%%%%%%%%%%%%%

The simplest possible approach to the $C$-AF phase is to assume that 
the coupling to the lattice dominates and stabilizes the $G$-AO order, 
as shown in Fig. \ref{fig:claso}(b). Such a robust OO would lead to 
classical values of orbital correlations in $G$-type phases, with
$\langle\vec\tau_i\cdot\vec\tau_j+\frac{1}{4}\rangle^{(c)}=0$, and 
$\langle\tau_i^y\tau_j^y\rangle^{(c)}=0$. However, it was recently 
pointed out\cite{Ole05} that this situation does not occur in LaVO$_3$, 
and instead one has to consider fluctuating orbitals. The exchange 
constants in the $C$-AF phase, $J_c^C$ and $J_{ab}^C$, can be found from 
the orbital excitations in the 1D disordered orbital chain, and we 
provide analytic expressions to evaluate them in Sec. \ref{sec:ow}. 
They allow one to determine the (weak) OO parameter 
$\langle\tau^z\rangle$ and the intersite orbital correlations which 
appear in Eqs. (\ref{jcd2}) and (\ref{jabd2}). Here we present the 
result of the numerical calculation, see Fig. \ref{fig:ex}(b). The FM 
exchange $J_c^C$ is {\it finite\/} already at $\eta=0$ due to the $a/b$ 
orbital fluctuations,\cite{Kha01} and is further enhanced by increasing 
splitting between the high-spin and low-spin excitations when Hund's 
exchange $\eta$ increases. At the same time, the AF exchange interaction 
$J_{ab}^C$ in $ab$ planes decreases.

Finally, we consider $G$-AF phase realized in YVO$_3$ at low temperature 
$T<T_{N2}$. A classical state with robust $C$-AO order has been proposed 
for this phase,\cite{Bla01} as shown in Fig. \ref{fig:claso}(c). 
We have verified that the quantum corrections to the OO 
parameter $\langle\tau^z\rangle$ are indeed negligible by considering 
the orbital waves for such a classical $C$-AO phase, see Sec. 
\ref{sec:ow}, so one finds indeed rather simple expressions for the AF 
exchange constants along the $c$ axis and in $ab$ planes:
\begin{eqnarray}
\label{jcgaf}
J_c^{G}&=&\frac{1}{4}J(1-\eta r_3),         \\
\label{jagaf}
J_{ab}^{G}&=&\frac{1}{4}J[1-\eta(r_1+r_3)].
\end{eqnarray}
Both above coupling constants decrease with increasing Hund's exchange, 
and the anisotropy between $J_c$ and $J_{ab}$ is gradually enhanced 
(Fig. \ref{fig:ex}).

%%%%%%%%%%%%%%%%%%%%%%%%%%%%%%%%%%%%%%%%%%%%%%%%%%%%%%%%%%%%%%%%%%%%%%%%
%%                         Spin wave excitations   
%%%%%%%%%%%%%%%%%%%%%%%%%%%%%%%%%%%%%%%%%%%%%%%%%%%%%%%%%%%%%%%%%%%%%%%%
\subsection{ Spin wave excitations }
\label{sec:sw}

The spin waves in different phases can be derived using the linear spin 
wave (LSW) theory.\cite{Mat81,Tak89} In the present case of $S=1$ spins 
this approach gives reliable results also for the OVB phase, in 
constrast to the linear orbital wave (LOW) theory for $\tau=1/2$ 
pseudospins which cannot be applied to the OVB phase as the $a$ and $b$
orbitals are there disordered. For the AF phases with two (or four) 
sublattices and the classical AF order $\langle S^z_i\rangle=\pm S$ 
considered here, we first rotate the spin operators on the sites 
occupied by down spins (with $\langle S^z_i\rangle=-S$) by angle $\pi$ 
with respect to spin $x$ axis, which leads to the canonical 
transformation:
\begin{equation}
\label{rota}
S_i^{\pm}\Rightarrow -S_i^{\pm}, \hskip .7cm S_i^z\Rightarrow -S_i^z.
\end{equation}
Next we write the equations of motion for the spin operators and apply 
standard Holstein-Primakoff transformation\cite{Mat81} from spin 
operators to boson operators (here $S=1$), 
\begin{equation}
\label{bosons}
S_i^+\simeq \sqrt{2S} a_i^{},        \hskip .5cm 
S_i^-\simeq \sqrt{2S} a_i^{\dagger}, \hskip .5cm
S_i^z = S-a_i^{\dagger}a_i^{}.
\end{equation}
The respective boson problem is easily diagonalized by employing first 
the Fourier transformation and next a Bogoliubov transformation in the 
momentum space ${\bf k}$. 

%%%%%%%%%%%%%%%%%%%%%%%%%%%%%%%%%%%%%%%%%%%%%%%%%%%%%%%%%%%%%%%%%%%%%%%%
%%                             Figure 5
%%%%%%%%%%%%%%%%%%%%%%%%%%%%%%%%%%%%%%%%%%%%%%%%%%%%%%%%%%%%%%%%%%%%%%%%
\begin{figure}[t!]
\includegraphics[width=7cm]{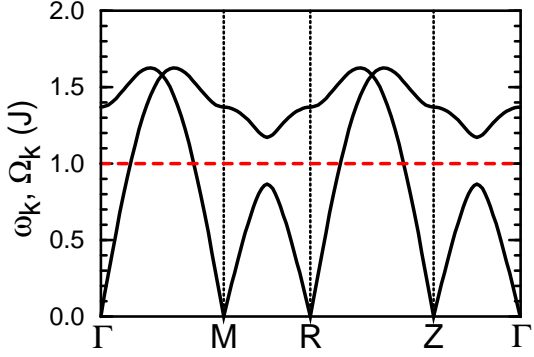}
\caption{(Color online)
Spin-wave dispersion $\omega_{\bf k}$ (full lines) and orbital  
triplet excitation energy $\Omega_{\bf k}$ (dashed line), as 
obtained for the OVB phase along high symmetry directions in the 
Brillouin zone at $\eta=0$ and $V_c=0$. For the present parameters 
one finds the following values of spin exchange constants:
$J_{ab}=0.3125J$, $J_{c1}=-0.250J$ and $J_{c2}=0.125J$. 
The high symmetry points are: 
$\Gamma=(0,0,0)$, $M=(\pi,\pi,0)$, $R=(\pi,\pi,\pi)$, $Z=(0,0,\pi)$. 
} 
\label{fig:ovb}
\end{figure}

Following the above procedure, one finds the spin-wave dispersion in 
the OVB phase,
\begin{eqnarray}
\label{swo}
\omega_{O\pm}({\bf k})&=&\Big\{\big[4J_{ab}+|J_{c1}|+J_{c2}\big]^2
-\big[4J_{ab}\gamma({\bf k})                               \nonumber \\
&\pm&\{(|J_{c1}|+J_{c2})^2
           -4J_{c1}J_{c2}\cos^2\!k_z\}^{1/2}\big]\Big\}^{1/2},\nonumber \\     
\end{eqnarray}
where the dispersion due to the AF exchange $J_{ab}$ coupling in $ab$ 
planes depends on the two-dimensional structure function,
\begin{equation}
\label{gama}
\gamma({\bf k}) = \frac{1}{2}(\cos k_x+\cos k_y).
\end{equation}
The two branches of $\omega_{O\pm}({\bf k})$ follow from the alternating 
FM $J_{c1}<0$ Eq. (\ref{djc1}) and AF $J_{c2}>0$ Eq. (\ref{djc2}) 
exchange interactions along the $c$ axis in a dimerized OVB state, shown 
in Fig. \ref{fig:claso}(a). For the case of $\eta=0$ the spin
waves extend up to $\sim 1.62J$ (Fig. \ref{fig:ovb}).

\begin{figure}[t!]
\includegraphics[width=7cm]{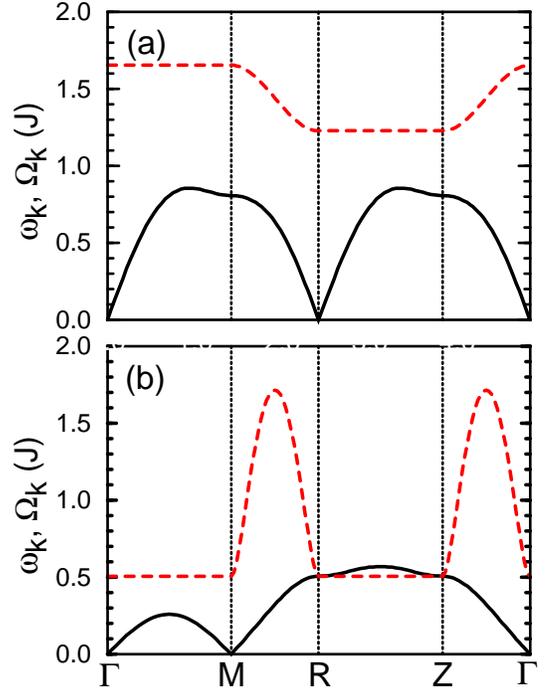}
\caption{(Color online)
Spin-wave dispersions $\omega_{\bf k}$ (full lines) as obtained in the 
LSW theory for the parameters motivated by experiment\cite{Ulr03} 
(for $J=40$ meV):
(a) $G$-AF phase with $J_{ab}=0.1425J$ and $J_c=-0.1425J$; 
(b) $C$-AF phase with $J_{ab}=0.0650J$ and $J_c= 0.0775J$.
Orbital excitations $\Omega_{\bf k}$ (dashed lines) were obtained 
within the LOW theory for the $G$-AF phase (a), and for a disordered 
orbital state in $C$-AF phase (b). 
Other parameters: $\eta=0.13$, $V_a=0.3J$, $V_c=0.84J$. 
High symmetry points as in Fig. \ref{fig:ovb}.
} 
\label{fig:cgaf}
\end{figure}

For the $G$-AF phase one finds the spin waves which depend on the 
(weakly anisotropic) AF exchange interactions given by Eqs. 
(\ref{jcgaf}) and (\ref{jagaf}),
\begin{equation}
\label{swg}
\omega_G({\bf k})=2\Big\{\big(2J_{ab}+J_c\big)^2
-\big(2J_{ab}\gamma_{\bf k}+J_c\cos k_z\big)^2\Big\}^{1/2}.
\end{equation}
For the numerical evaluation we ignored weak anisotropy of the magnetic 
exchange constants which follows from the spin-orbital model, and 
adopted the experimental isotropic parameters $J_{ab}=J_c=5.7$ meV,
i.e., $J_{ab}=J_c=0.1425J$ for $J=40$ meV.\cite{Ulr03} These parameters 
are somewhat lower than those which would result from Eqs. (\ref{jcgaf}) 
and (\ref{jagaf}) for the present value of $J$ at $\eta=0.13$, and give 
the width of the magnon dispersion close to $0.85J$, 
see Fig. \ref{fig:cgaf}(a).

%%%%%%%%%%%%%%%%%%%%%%%%%%%%%%%%%%%%%%%%%%%%%%%%%%%%%%%%%%%%%%%%%%%%%%%%
%%                             Figure 6
%%%%%%%%%%%%%%%%%%%%%%%%%%%%%%%%%%%%%%%%%%%%%%%%%%%%%%%%%%%%%%%%%%%%%%%%

Finally, we consider the $C$-AF phase with uniform FM interactions  
$J_c$ for which one finds the spin-wave dispersion (for more 
details see Appendix \ref{sec:quantum})
\begin{equation}
\label{swc0}
\omega_C^{(0)}({\bf k})=2\Big\{\big[2J_{ab}+|J_c|(1+\cos k_z)\big]^2
-\big(2J_{ab}\gamma_{\bf k}\big)^2\Big\}^{1/2}.
\end{equation}
This result corresponds to an idealized structure when the observed 
alternation of stronger and weaker FM interactions along the $c$ axis 
(see Sec. \ref{sec:red}) is
ignored. Taking again the experimental exchange constants:\cite{Ulr03} 
$J_{ab}=2.6$ meV and $J_c=3.1$ meV, i.e., $J_{ab}=0.065J$ and 
$J_c=0.077J$ for $J=40$ meV, one finds that the spin wave spectrum  
extends up to $0.57J$, see Fig. \ref{fig:cgaf}(b). Therefore, due to 
the observed strong reduction of the exchange interactions,\cite{Ulr03} 
the overall width of the magnon band is {\it lower\/} in $C$-AF phase, 
while the theory predicts\cite{Rac02} here a {\it wider\/} magnon band 
for rather similar exchange interactions in both $C$-AF and $G$-AF 
phase, as they follow from Eq. (\ref{HJ}). 

%%%%%%%%%%%%%%%%%%%%%%%%%%%%%%%%%%%%%%%%%%%%%%%%%%%%%%%%%%%%%%%%%%%%%%%%
%%                         Orbital excitations   
%%%%%%%%%%%%%%%%%%%%%%%%%%%%%%%%%%%%%%%%%%%%%%%%%%%%%%%%%%%%%%%%%%%%%%%%
\subsection{ Orbital excitations }
\label{sec:ow}

In contrast to spins which show particular types of long range order
in various phases of Fig. \ref{fig:claso}, the orbitals are in the first 
instance disordered due to a robust tendency towards strong 1D 
fluctuations of $a$ and $b$ orbitals along $c$ axis.\cite{Sir03,Miy05} 
This may change, however, when lattice distortions [which induce 
intersite orbital interactions in ${\cal H}_{\rm orb}$ in Eq. 
(\ref{model})] contribute and stabilize a particular type of AO 
order. Therefore, one has to 
employ different approaches to determine the orbital excitations ---
they depend on the parameter regime and on the underlying orbital phase. 

First, in the OVB phase the orbitals are entirely disordered, and one 
has only the short range order of orbital singlets along $c$ axis which 
imposes the dimerized magnetic phase, see Fig. \ref{fig:claso}(a). 
Under these circumstances, one finds an orbital triplet excitation for 
each singlet bond along $c$ axis which supports local FM spin order, 
\begin{equation}
\label{owo}
\Omega_O({\bf k}) = J\Big(r_1+\frac{1}{4}v_c\Big).
\end{equation}
As these orbital excitations are local, they are dispersionless and
involve no further quantum correction to the energy $E_{\rm OVB}$ 
given by Eq. (\ref{eovb}). 

Second, although we will show below that $a$ and $b$ orbitals are to 
some extent disordered in $C$-AF phase, weak long range order survives 
in the relevant range of parameters near $\eta\sim 0.13$, so we may
start with a classical $G$-AO order at $T=0$, and make an expansion 
around this state using Gaussian fluctuations. In this approach one 
rotates first the orbital operators on the sites occupied by $b$ 
orbitals (down pseudospins) with $\langle\tau^z_i\rangle=-\frac{1}{2}$ 
by angle $\pi$ with respect to pseudospin $x$ axis, which leads to the
canonical transformation:
\begin{equation}
\label{rotatau}
\tau_i^{\pm}\Rightarrow -\tau_i^{\pm}, \hskip .7cm 
\tau_i^z    \Rightarrow -\tau_i^z.
\end{equation}
Next we introduce a similar expansion to that considered above for the 
spin operators,\cite{Kha01} and express the orbital operators in terms 
of the respective Holstein-Primakoff bosons $\{b_i,b_i^{\dagger}\}$,
\begin{equation}
\label{ow}
\tau_i^+\simeq b_i^{}, \hskip .5cm \tau_i^-\simeq b_i^{\dagger}, 
\hskip .5cm \tau_i^z = \frac{1}{2}-b_i^{\dagger}b_i^{}.
\end{equation}
Here we assumed a robust $G$-type OO ($G$-AO) state which 
may be used as a classical state to determine the orbital excitations 
by performing a Gaussian expansion around it. When only the leading 
terms are kept within the LOW theory,\cite{Bri99} 
one finds after the Fourier transformation and the subsequent 
Bogoliubov transformation the orbital-wave energy,
\begin{equation}
\label{owc0}
\Omega_C^{(0)}(k_z) =J\big\{\Delta^2+r_1^2\sin^2k_z\big\}^{1/2}.
\end{equation}
The spectrum has a gap at $k_z=0$ 
\begin{equation}
\label{owcgap}
\Delta=\left\{\big[\eta(r_1+r_3)+v_0\big]
              \big[2r_1+\eta(r_1+r_3)+v_0\big]\right\}^{1/2}
\end{equation}
at $k_z=0$, where 
\begin{equation}
\label{v0}
v_0=2v_a-v_c. 
\end{equation}
Note that the gap 
$\Delta$ depends on a linear combination of orbital interactions $v_0$ 
for the present form of Eq. (\ref{Horb}), so the interactions along $c$ 
axis and the ones in $ab$ planes partly compensate each other in Eq. 
(\ref{v0}). In fact, Eq. (\ref{owc0}) reproduces the earlier 
result obtained for $v_0=0$ in Ref. \onlinecite{Kha01}, but in general 
both types of orbital interactions originate from different distortions
and are thus independent from each other. 
The orbiton dispersion demonstrates that the present phase is stable 
at finite $\eta$ only as long as $\Delta>0$, i.e., in a range of 
$v_0>-\eta(r_1+r_3)$. 
The orbital-wave dispersion (\ref{owc0}) follows from the quantum 
fluctuations along the $c$ axis, and thus depends only on the $z$th 
momentum component $k_z$. We emphasize that the orbital 
excitations are typically at higher energy than the spin excitations 
as the orbital gap $\Delta$ is finite, see Fig. \ref{fig:cgaf}(b). 

Here we also give the values of the orbital correlations which enter
Eqs. (\ref{jcd2}) and (\ref{jabd2}). It is convenient to introduce 
the following integrated quantities: 
\begin{eqnarray}
\label{s1}
s_1&=&\frac{1}{2N}\sum_k\Big\{A-\Omega_C^{(0)}(k)\Big\},   \\
\label{s2}
s_2&=&\frac{1}{2N}\sum_k\left\{\frac{A}{\Omega_C^{(0)}(k)}-1\right\},
\end{eqnarray}
where $A=r_1+2\eta(r_1+r_3)\langle\tau^z\rangle$. The orbital 
correlations and the OO parameter,
\begin{equation}
\label{tz}
\langle\tau^z\rangle=\frac{1}{2}-s_2,
\end{equation}
are reduced by quantum fluctuations along the $c$ axis and
are determined self-consistently. One finds that weak OO 
appears at finite $\eta>0.08$ (at $\eta<0.08$ the orbitals are 
disordered and $\langle\tau^z\rangle=0$), and 
$\langle\tau^z\rangle\simeq 0.26$ for $\eta=0.13$, i.e., the OO is 
only about {\it half of the classical value\/} for the realistic 
parameters of cubic vanadates.\cite{Kha04} This demonstrates that the 
$\{a,b\}$ orbitals strongly fluctuate and the $G$-AO order in rather 
weak. Strong orbital fluctuations can be also verified by calculating 
the intersite orbital correlations: 
\begin{equation}
\label{ttcaf}
\langle\vec\tau_i\!\cdot\!\vec\tau_j\rangle=-\frac{1}{4}
-\frac{1}{r_1}\left[s_1+\eta(r_1+r_3)s_2\right].
\end{equation}
Indeed, one finds a rather low value of 
$\langle\vec\tau_i\!\cdot\!\vec\tau_j\rangle\simeq -0.428$ (not so far 
from the Bethe ansatz result -0.4431 for the AF Heisenberg chain), and 
the dominating contribution comes not from the static term 
$\langle\tau_i^z\tau_j^z\rangle=-\langle\tau^z\rangle^2\simeq -0.068$, 
but from the fluctuating part, 
$\langle\tau_i^x\tau_j^x+\tau_i^y\tau_j^y\rangle\simeq -0.36$.

Finally, the opposite situation is found in the $G$-AF phase, where 
structural distortions observed below $T_{N2}$ suggest that the $C$-AO 
order sets up. In this case the $a$ and $b$ orbitals repeat each other 
along the chains in $c$ direction, and alternate in $ab$ planes 
[Fig. \ref{fig:claso}(c)]. This robust $C$-type orbitally ordered state 
may be used to determine the orbital excitations employing the LOW theory.
\cite{Bri99} We used again a rotation of 'down' pseudospins as in Eq.
(\ref{rotatau}) in order to obtain a uniform ferro-orbital state, and 
next expressed the orbital operators in terms of the Holstein-Primakoff 
bosons $\{b_i,b_i^{\dagger}\}$, using Eqs. (\ref{ow}). By applying 
a similar procedure to that used above for the $G$-AO phase, i.e., 
keeping only bilinear terms in the leading LOW order, and employing 
subsequent Fourier and Bogoliubov transformations, this leads to the 
orbital waves in the $C$-AO phase, with dispersion
\begin{equation}
\label{owg}
\Omega_G({\bf k}) = J\big(\eta r_1\cos k_z+v_c+2v_a\big),
\end{equation}
characterized by a large gap of $\sim (V_c+2V_a)$, shown in Fig. 
\ref{fig:cgaf}(a). We emphasize that the interactions with the lattice 
are here of crucial importance and generate a large gap, while the 
orbital gap found in the $C$-AO phase follows predominantly from the 
superexchange interactions and is therefore typically much smaller than 
the one in the $G$-AO phase.\cite{Ish04}

%%%%%%%%%%%%%%%%%%%%%%%%%%%%%%%%%%%%%%%%%%%%%%%%%%%%%%%%%%%%%%%%%%%%%%%%
%%                           phase diagram
%%%%%%%%%%%%%%%%%%%%%%%%%%%%%%%%%%%%%%%%%%%%%%%%%%%%%%%%%%%%%%%%%%%%%%%%
\subsection{ Zero temperature phase diagram }
\label{sec:phd}

In order to investigate the relative stability of the magnetic phases 
shown schematically in  Fig. \ref{fig:claso}, one has to determine the 
quantum corrections due to magnetic and orbital excitations. The quantum 
corrections due to orbital fluctuations were already included in the 
energies of the OVB (\ref{eovb}) and $C$-AF (\ref{ecaf}) phases, 
where the orbital singlets along $c$ axis dominate and are responsible 
either for the orbital disordered state or for weak $G$-AO order, 
respectively. The quantum correction to the energy of the $G$-AF phase 
due to the almost dispersionless orbital waves Eq. (\ref{owg}) is rather 
small and will be neglected below.\cite{noteowg} 

%%%%%%%%%%%%%%%%%%%%%%%%%%%%%%%%%%%%%%%%%%%%%%%%%%%%%%%%%%%%%%%%%%%%%%%%
%%                             Figure 7
%%%%%%%%%%%%%%%%%%%%%%%%%%%%%%%%%%%%%%%%%%%%%%%%%%%%%%%%%%%%%%%%%%%%%%%%
\begin{figure}[t!]
\includegraphics[width=7cm]{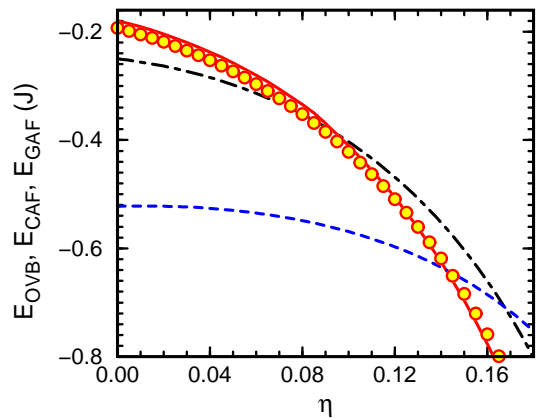}
\caption{(Color online)
Energies of different phases for increasing Hund's exchange $\eta$: 
$C$-AF phase (solid line), OVB phase (dashed-dotted line), and $G$-AF 
phase (long-dashed line). The energy of the $G$-AF phase is shown for 
$2V_a=V_c=0.45J$ (the other energies do not depend on $V_c$). 
A constant energy term $-2J$ was neglected in all phases. The circles 
show for a comparison the energy obtained for decoupled FM chains along 
$c$ axis, with orbital correlations described by the 1D pseudospin 
Heisenberg model. 
} 
\label{fig:ene}
\end{figure}

The remaining quantum corrections to the classical energy of the N\'eel 
state due to spin excitations can be found using the standard approach 
of the LSW theory. At $T=0$ the total energy, 
\begin{equation}
\label{e}
E_M = E_M^{(0)}-\delta E_M,
\end{equation}
is lowered by the quantum fluctuation contribution\cite{Rac02}
\begin{equation}
\label{deltae}
\delta E_M = 2J_{ab}+|J_c|
-\frac{1}{2(2\pi)^3}\int d^3{\bf k}\;\;\omega_M({\bf k}),
\end{equation}
where label $M=0,C,G$ stands for a given magnetic phase considered 
here, either OVB, or $C$-AF, or $G$-AF;
while $\omega_M({\bf k})$ in Eq. (\ref{deltae})
is the spin wave dispersion in this phase. We have 
evaluated quantum corrections using Eq. (\ref{deltae}) for all three 
magnetic phases: OVB, $C$-AF and $G$-AF. It is instructive to investigate 
first the energy dependence on Hund's exchange interaction, shown in Fig. 
\ref{fig:ene}. As the quantum corrections which result from spin 
excitations are similar for all three AF phases, the qualitative picture 
obtained with these corrections and presented in Sec. \ref{sec:claso} is 
confirmed: the $C$-AF is stable in a range of realistic values of Hund's 
exchange $\eta\sim 0.13$ for small orbital interaction parameter $V_c$, 
while increasing this interaction results in a transition to the $G$-AF 
phase, where the magnetic energy is gained on all the bonds after 
the orbitals have reoriented to the $C$-AO order, see 
Fig. \ref{fig:claso}. 

A transition from the OVB phase to the $C$-AF one under increasing 
$\eta$ is rather intricate. At small values of $\eta$ when the OVB 
phase is still stable, the competing phase with $C$-AF spin order is 
the orbital disordered phase, as the orbital superexchange interactions 
in $ab$ planes are so weak (and the orbital interactions cancel each 
other on the mean field level for $2V_a=V_c$) that the 1D pseudospin 
interaction along the $c$ axis dominates\cite{Sir03} the behavior of 
the orbital chain (see Fig. \ref{fig:ene}). However, at $\eta\sim 0.13$ 
one finds that weak $G$-AO order is stabilized by Ising orbital 
interactions along the bonds in $ab$ planes. However, the orbital 
fluctuations are still very strong in this state as described in Sec. 
\ref{sec:ow}. Of course, the $G$-AO order could be further stabilized 
and become of more classical character\cite{Bla01} when $2V_a>V_c$, 
but this picture of the $C$-AF phase contradicts recent experiments.
\cite{Miy06} 

%%%%%%%%%%%%%%%%%%%%%%%%%%%%%%%%%%%%%%%%%%%%%%%%%%%%%%%%%%%%%%%%%%%%%%%%
%%                             Figure 8
%%%%%%%%%%%%%%%%%%%%%%%%%%%%%%%%%%%%%%%%%%%%%%%%%%%%%%%%%%%%%%%%%%%%%%%%
\begin{figure}[t!]
\includegraphics[width=7cm]{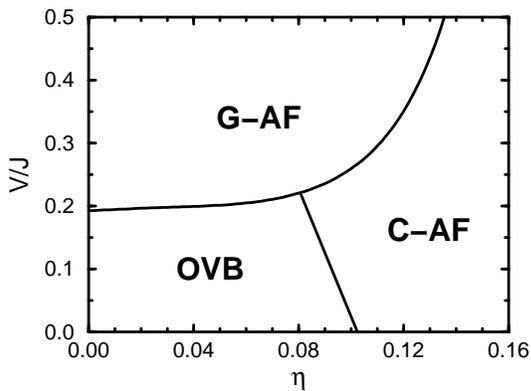}
\caption{
Mean-field phase diagram of the spin-orbital model (\ref{model}) as 
obtained for cubic vanadates in the $(\eta,V)$ plane at $T=0$ for 
$V_c=2V_a=2V$. At the spectroscopic
value of $\eta\simeq 0.13$ two phases are possible: $C$-AF phase 
(for $V<V_0$) and $G$-AF phase (for $V>V_0$), with $V_0\simeq 0.43J$; 
these two AF states are observed at low temperature in LaVO$_3$ 
($C$-AF) and in YVO$_3$ ($G$-AF), respectively. 
} 
\label{fig:phd}
\end{figure}

By comparing energies of all three magnetic phases at $T=0$ one finds
the phase diagram of Fig. \ref{fig:phd}. To simplify the discussion we
have adopted here the parametrization $V_c=2V_a=2V$. In fact, one 
expects that the GdFeO$_3$-like distortions are responsible for stronger 
orbital interactions along $c$ axis, and the parameter $V_c$ plays a 
more important role (than $V_a$) in stabilizing the $C$-AO order which 
supports the $G$-AF spin order. The OVB phase is stable for small values 
of $\eta$ and $V$, while for sufficiently large $V$ the $G$-AF phase 
takes over. At larger values of $\eta$ two AF phases observed in the 
cubic vanadates,\cite{Miy06} $C$-AF and $G$-AF phase, compete with each 
other. The range of stability of the $C$-AF phase increases with 
increasing $\eta$ as the FM interaction along $c$ axis is then enhanced, 
see Fig. \ref{fig:ex}(b). In contrast, both AF exchange interactions in 
the $G$-AF phase are reduced, so this phase has to be stabilized by 
larger orbital interaction $V$.

%%%%%%%%%%%%%%%%%%%%%%%%%%%%%%%%%%%%%%%%%%%%%%%%%%%%%%%%%%%%%%%%%%%%%%%%
%%                        Scenario for YVO$_3$ 
%%%%%%%%%%%%%%%%%%%%%%%%%%%%%%%%%%%%%%%%%%%%%%%%%%%%%%%%%%%%%%%%%%%%%%%%
\section{ Scenario for YVO$_3$ }
\label{sec:scenario}

%%%%%%%%%%%%%%%%%%%%%%%%%%%%%%%%%%%%%%%%%%%%%%%%%%%%%%%%%%%%%%%%%%%%%%%%
%%                              Peierls
%%%%%%%%%%%%%%%%%%%%%%%%%%%%%%%%%%%%%%%%%%%%%%%%%%%%%%%%%%%%%%%%%%%%%%%%
\subsection{ Peierls orbital dimerization }
\label{sec:dim}

Before we address the experimental situation in YVO$_3$, we demonstrate
an intrinsic instability of the 1D spin-orbital chain towards 
dimerization.\cite{Sir03} In contrast to the 1D Heisenberg 
antiferromagnet with fixed exchange interactions on each bond 
$\langle i,i+1\rangle$, the orbital interaction in the present case are 
[in the leading order, see Eq. (\ref{orbj})] given by
\begin{equation}
\label{jo}
J_{\rm orb}(i,i+1) = \frac{1}{2}(1+2\eta r_1)\,
\Big\langle{\vec S}_i\cdot{\vec S}_{i+1}+1\Big\rangle~,
\end{equation}
i.e., for each bond the orbital interaction is tuned by the spin 
correlation function on this bond. While at temperature $T=0$ the spins 
are (almost) fully polarized and 
$\langle{\vec S}_i\cdot{\vec S}_{i+1}\rangle\simeq 0.96$,\cite{Rac02}
the spin correlations could in principle alternate between stronger and 
weaker FM bonds at finite temperature $T>0$, and then the orbital 
interaction would be modulated as follows
\begin{equation}
\label{jopm}
J_{\rm orb}(i,i+1) = J_o(1\pm\delta_o),
\end{equation}
between even and odd bonds.
Note that $J_o$ stands here for the average value that will gradually 
decrease with increasing temperature. This additional temperature 
dependence complicates somewhat the picture of the $C$-AF phase.

\begin{figure}[t!]
\includegraphics[width=7cm]{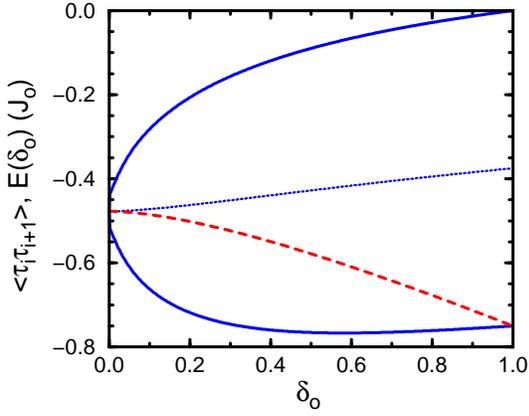}
\caption{(Color online)
Orbital correlation functions 
$\langle{\vec\tau}_i\cdot {\vec\tau}_{i+1}\rangle$ at even and odd 
bonds along a dimerized 1D chain, as obtained within the XY model for 
increasing anisotropy $\delta_o$ in the exchange constants, see
Eq. (\ref{jopm}). The energy $E(\delta_o)$ (dashed line), obtained 
using spinless fermions Eq. (\ref{edim}), decreases with
increasing $\delta_o$, while the average energy in an orbital chain 
with the same exchange interaction $J_o$ at each bond would increase 
(dotted line). 
} 
\label{fig:dimero}
\end{figure}

Assuming the alternating orbital interactions (\ref{jopm}) and 
performing the transformation to fermions for the corresponding XY 
model in the orbital sector, one finds the following spinless fermion 
problem using the Jordan-Wigner transformation\cite{JW} 
\begin{equation}
\label{HXY}
{\cal H}_{XY}(\delta_o) = \frac{1}{2}J_o\sum_i(1\pm\delta_o)
\big(f_i^{\dagger}f_{i+1}^{}+f_{i+1}^{\dagger}f_i^{}\big).
\end{equation}
The diagonalization of Hamiltonian (\ref{HXY}) gives the energy 
spectrum of a dimerized fermionic chain,
\begin{equation}
\label{ek}
\varepsilon_{\pm}(k) = \pm \sqrt{\cos^2 k+\delta^2\sin^2 k},
\end{equation}
and the total energy at $T=0$:
\begin{equation}
\label{edim}
E(\delta_o) = -J_o\frac{3}{2\pi}\int_0^{\pi/2}
\;dk\; \varepsilon_{-}(k).
\end{equation}
The energy $-0.4776J$ obtained from Eq. (\ref{edim}) at $\delta_o=0$
is slighly lower than the Bethe ansatz result ($-0.4431J$), while at 
$\delta_o=1$ the exact result found for the orbital singlets on every 
second bond is rigorously reproduced. Therefore, Eq. (\ref{edim}) may 
be considered to be a reasonable interpolation formula which allows one 
to investigate the dimerized orbital chain in the entire regime of 
$\delta_o$. While an average value of the orbital correlation function
$\langle{\vec\tau}_i\cdot{\vec\tau}_{i+1}\rangle$ {\it increases\/}
with $\delta_o$, the chain with a constant exchange interaction cannot 
dimerize by itself. In contrast, the energy $E(\delta_o)$ indeed 
{\it decreases\/} when the alternation of the orbital 
interactions (\ref{jopm}) is allowed, so the chain does have 
a tendency to dimerize (Fig. \ref{fig:dimero}). 

%%%%%%%%%%%%%%%%%%%%%%%%%%%%%%%%%%%%%%%%%%%%%%%%%%%%%%%%%%%%%%%%%%%%%%%%
%%                             Figure 9
%%%%%%%%%%%%%%%%%%%%%%%%%%%%%%%%%%%%%%%%%%%%%%%%%%%%%%%%%%%%%%%%%%%%%%%%

\begin{figure}[t!]
\includegraphics[width=7cm]{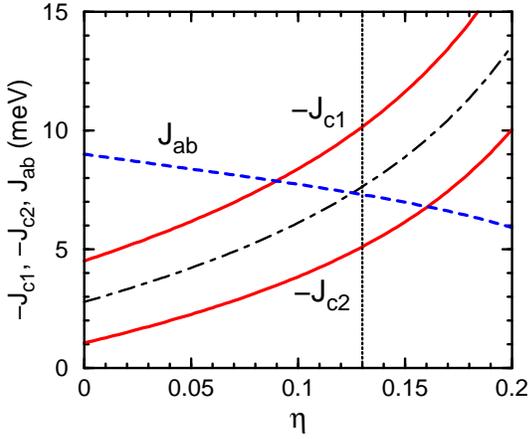}
\caption{(Color online)
Exchange constants $J_{ab}$ (dashed line), $J_{c1}$ and $J_{c2}$ 
(solid lines), all in meV, as obtained for the idealized $C$-AF phase 
with condensed $xy$ orbitals ($n_c=1$), Eqs. (\ref{frozen}), and orbital 
disordered state along $c$ axis with dimerized orbital correlations.
Vertical line indicates the value of $\eta=0.13$ estimated from the 
atomic data\cite{Zaa90} and from the optical data\cite{Kha04} for 
LaVO$_3$. Parameters: $J=30$ meV, 
$\langle{\vec\tau}_i\cdot {\vec\tau}_{i+1}\rangle=-0.4431$, 
$\delta_{\tau}=0.12$ (\ref{deltat}).
} 
\label{fig:jcaf}
\end{figure}

It is quite remarkable that already a weak anisotropy $\delta_o$ in
orbital interactions is sufficient to give rather different orbital 
correlations $\langle{\vec\tau}_i\cdot{\vec\tau}_{i+1}\rangle$ on 
even/odd bonds. These different orbital correlations can trigger the 
alternation in the spin correlation functions, and in this way the 
dimerized state could be a self-consistent solution of the spin-orbital 
problem at finite temperature. We emphasize that even a relatively 
small anisotropy $\delta_{\tau}=0.12$ in the orbital correlations,
\begin{equation}
\label{deltat}
\delta_{\tau}=|\langle{\vec\tau}_ i   \cdot{\vec\tau}_{i+1}\rangle
              -\langle{\vec\tau}_{i+1}\cdot{\vec\tau}_{i+2}\rangle|,
\end{equation}
is already sufficient to generate considerable anisotropy in the 
magnetic exchange constants $J_{c1}$ and $J_{c2}$ along $c$ axis (Fig. 
\ref{fig:jcaf}). The exchange constants of Fig.  \ref{fig:jcaf} were 
obtained with $J=30$ meV --- this reduction of the energy scale by a 
semiempirical factor of 0.75 from that given by the analysis of the 
optical spectrum\cite{Kha04} was necessary as otherwise the model 
(\ref{model}) would predict too large exchange constants for the 
$G$-AF phase. Furthermore, we note that the above anisotropy 
$\delta_{\tau}$ is obtained already with $\delta_o=0.017$ when the 
mapping to the fermion problem (\ref{HXY}) is used (Fig. 
\ref{fig:dimero}). Of course, this problem requires a self-consistent 
solution at finite temperature as we discuss in Sec. \ref{sec:red}.

%%%%%%%%%%%%%%%%%%%%%%%%%%%%%%%%%%%%%%%%%%%%%%%%%%%%%%%%%%%%%%%%%%%%%%%%
%%                             Figure 10
%%%%%%%%%%%%%%%%%%%%%%%%%%%%%%%%%%%%%%%%%%%%%%%%%%%%%%%%%%%%%%%%%%%%%%%%

%%%%%%%%%%%%%%%%%%%%%%%%%%%%%%%%%%%%%%%%%%%%%%%%%%%%%%%%%%%%%%%%%%%%%%%%
%%                           orbital fluctuations 
%%%%%%%%%%%%%%%%%%%%%%%%%%%%%%%%%%%%%%%%%%%%%%%%%%%%%%%%%%%%%%%%%%%%%%%%
\subsection{ Reduction of exchange constants by orbital fluctuations }
\label{sec:red}

Although the value of $J\sim 40$ meV deduced from the neutron 
scattering data\cite{Ulr03} for YVO$_3$ gives a consistent description  
of the temperature dependence of the optical spectral weight for the 
high-spin excitations along $c$ axis in LaVO$_3$, there is a fundamental 
problem concerning the size of magnetic exchange constants, particularly 
in the exotic $C$-AF phase of YVO$_3$, stable in the intermediate 
temperature range $T_{N2}<T<T_{N1}$. First of all, the calculations 
performed using the mean-field approach and assuming rigid 
OO (see Fig. \ref{fig:claso}), as in Ref. \onlinecite{Ole05}, predict 
too large exchange constants in both phases when $J=40$ meV is assumed. 
In fact, for the $G$-AF phase one finds then the values of both $J_c$ 
and $J_{ab}$ being larger by $\sim 25$\% than the
respective experimental values of Ulrich {\it et al.\/}.\cite{Ulr03}
Moreover, in experiment one finds an (almost) isotropic $G$-AF phase
with $J_c=J_{ab}$, while the present model predicts (except at small 
$\eta<0.10$) an anisotropy between $c$ axis and $ab$ planes, with 
$J_c>J_{ab}$, see Eqs. (\ref{jcgaf}) and (\ref{jagaf}). 
This suggests that already for the $G$-AF phase some 'dynamical' 
reduction mechanism of the magnetic exchange constants is at work, 
which we simulate by reducing the superexchange energy scale down to 
$J\sim 30$ meV. Indeed, taking an average value of the magnetic 
exchange constants over three cubic directions we arrive then at the 
experimental result $J_c=J_{ab}\sim 5.7$ meV.

While the above procedure could be still considered as a fair agreement
between the theoretical model and experiment, it is surprising that 
the magnetic exchange constants in the $C$-AF phase cannot be obtained 
from the model using the same parameters. In fact, the values of $J_c$     
and $J_{ab}$ shown in Fig. \ref{fig:jcaf} for $\eta=0.13$ are by  
almost a factor of 2 larger than those deduced from the neutron 
scattering data at 85 K.\cite{Ulr03} This strongly suggests that some
of the assumptions used so far to derive the values of $J_c^C$ and 
$J_{ab}^C$ from Eqs. (\ref{jcd2}) and (\ref{jabd2}) have to be
reconsidered. 

%%%%%%%%%%%%%%%%%%%%%%%%%%%%%%%%%%%%%%%%%%%%%%%%%%%%%%%%%%%%%%%%%%%%%%%%
%%                             Figure 11
%%%%%%%%%%%%%%%%%%%%%%%%%%%%%%%%%%%%%%%%%%%%%%%%%%%%%%%%%%%%%%%%%%%%%%%%
\begin{figure}[t!]
\includegraphics[width=7cm]{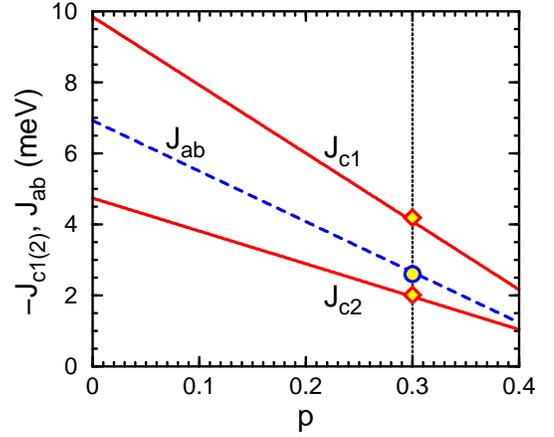}
\caption{(Color online)
Reduction of exchange constants ${\cal J}_{ab}$ (dashed line), 
${\cal J}_{c1}$ and 
${\cal J}_{c2}$ (solid lines), all in meV, 
in the dimerized $C$-AF phase due to orbital bond fluctuations between 
FM and AF bonds, as given by Eqs. (\ref{Jcp1})--(\ref{Jabp}).
Orbital disordered state along $c$ axis is assumed at $p=0$.
Experimental values of exchange constants found for the $C$-AF phase of 
YVO$_3$,\cite{Ulr03} shown by circle 
($J_{ab}\simeq 2.6$ meV) and by diamonds 
($J_{c1}\simeq 4.2$ meV and $J_{c2}\simeq 2.0$ meV), 
are nearly reproduced for moderate fluctuations with $p=0.30$
(vertical dashed line).
Parameters: $J=30$ meV, $\eta=0.13$, $\delta_s=0.35$.  
} 
\label{fig:jplaq}
\end{figure}

One of the most puzzling experimental features in YVO$_3$ is the nature 
of the structural transition at $T_s$, which removes the orbital 
degeneracy and induces the splitting $\Delta$ between the $xy$ orbitals 
and the $yz/zx$ doublet, see Fig. \ref{fig:artist}. We anticipate that 
this splitting is not large enough to impose strict freezing of charge 
in $xy$ orbitals. Thus we expect that some orbital fluctuations should 
still be present in the intermediate temperature regime $T_{N2}<T<T_s$
before the orbitals undergo the transition into the $C$-AO phase 
(supporting $G$-AF spin order) below $T_{N2}$, as shown in Fig. 
\ref{fig:claso}(c). 

Qualitatively,
we illustrate the consequences of orbital fluctuations on the magnetic 
exchange constants by considering a plaquette which includes two bonds 
along $c$ axis and two other bonds along either $a$ or $b$ axis. If the
$c$ orbitals are occupied at each site, and $a/b$ orbitals fluctuate,
a representative state of such a plaquette contains 4 electrons in $c$ 
orbitals, and 2 in each of two other states, $a$ and $b$. The effective 
superexchange Hamiltonian (\ref{HJ}) contains the terms with double 
orbital excitations on the bonds, $\propto\tau_i^\pm\tau_j^\pm$, see Eq. 
(\ref{exo}). Such terms on the bonds along $a$ (or $b$) axis generate 
$a/b$ orbital configurations on each site $i$ and $j$. Only one of these
two orbitals ($a$ or $b$) is active along this particular bond, and it resembles 
the bond along $c$ axis before the orbital fluctuation took place. As
a result, such fluctuations lead to (locally) FM contributions in the  
$ab$ planes, and to (locally) AF contributions along $c$ axis --- both
of them will reduce the actual values of $J_c^C$ and $J_{ab}^C$ exchange
constants.  

Following the above idea, we introduce {\it effective} magnetic exchange 
constants, 
\begin{eqnarray}
\label{Jcp1}
{\cal J}_{c1}^C&=&(1+\delta_s)\big[(1-p)J_c^C(0)+pJ_{ab}^C(0)\big],  \\
\label{Jcp2}
{\cal J}_{c2}^C&=&(1-\delta_s)\big[(1-p)J_c^C(0)+pJ_{ab}^C(0)\big],  \\
\label{Jabp}
{\cal J}_{ab}^C&=&(1-p)J_{ab}^C(0)+   p J_c^C(0),
\end{eqnarray}
as a superposition of two contributions obtained for the undimerized 
state without $xy$ orbital fluctuations (for $n_c=1$), $J_c^C(0)$ and 
$J_{ab}^C(0)$, calculated as described in Sec. \ref{sec:allj}. 
The probabilities $(1-p)$ and $p$ refer to the initial state with $c$ 
orbitals occupied ($n_c=1$), and to the configuration with flipped 
orbitals after the plaquette fluctuation has occurred ($n_c=0$), 
respectively. The result of the numerical calculation for the usual 
parameters shows that one arrives almost at experimental values of the 
magnetic exchange constants when moderate orbital fluctuations with 
$p=0.30$ considerably reduce the exchange constants 
(see Fig. \ref{fig:jplaq}). For the experimental anisotropy $\delta_s$ 
one finds large alternation of the FM exchange constants along $c$ 
axis with respect to the average value, 
\begin{equation}
\label{Jcp}
{\cal J}_{c}^C=(1-p)J_c^C(0)+pJ_{ab}^C(0).
\end{equation}

%%%%%%%%%%%%%%%%%%%%%%%%%%%%%%%%%%%%%%%%%%%%%%%%%%%%%%%%%%%%%%%%%%%%%%%%
%%                             Figure 12
%%%%%%%%%%%%%%%%%%%%%%%%%%%%%%%%%%%%%%%%%%%%%%%%%%%%%%%%%%%%%%%%%%%%%%%%
\begin{figure}[t!]
\includegraphics[width=8cm]{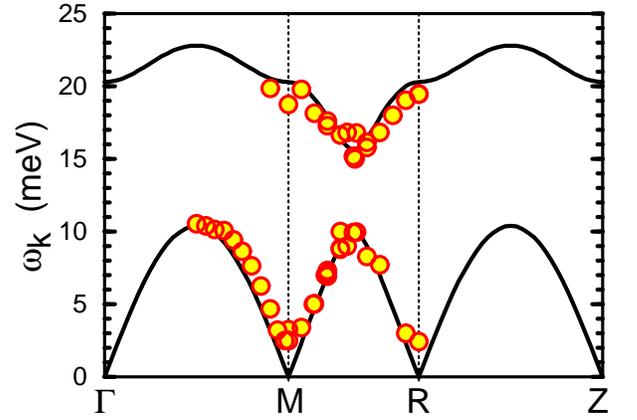}
\caption{(Color online)
Spin-wave dispersions $\omega_{\bf k}$ (full lines) as obtained 
in the LSW theory along the representative directions in the 
Brillouin zone for the dimerized $C$-AF phase with experimental 
exchange constants:\cite{Ulr03} 
$J_{ab}=2.6$ meV, $J_c=3.1(1\pm\delta_s)$ meV and $\delta_s=0.35$.
These interactions are obtained by considering plaquette fluctuations 
of spin exchange interactions as described in the text (see also Fig.
\ref{fig:jplaq}). 
Parameters: $J=30$ meV, $\eta=0.13$, $\delta_s=0.35$, $p=0.30$.
The experimental points of Ref. \onlinecite{Ulr03} measured by neutron 
scattering at $T=85$ K are reproduced by circles 
(the effective linewidths are not shown). The high symmetry points are: 
$\Gamma=(0,0,0)$, $M=(\pi,\pi,0)$, $R=(\pi,\pi,\pi)$, $Z=(0,0,\pi)$.
}
\label{fig:swcafd}
\end{figure}

Next, we analyze the spin excitations in the dimerized $C$-AF phase in
order to calculate the spin correlations, the quantum fluctuation 
correction to the ground state energy (see Appendix \ref{sec:quantum}), 
as well as the free energy at finite temperature, see Sec. 
\ref{sec:trans}. The effective spin Hamiltonian for this phase is given 
as follows:  
\begin{eqnarray}
\label{hcafd}
{\cal H}_s&=&
   {\cal J}_{c}^C(1+\delta_s)\sum_{\langle 2i,2i+1\rangle\parallel c}
      {\vec S}_{2i}\cdot{\vec S}_{2i+1}                     \nonumber \\
   &+&{\cal J}_{c}^C(1-\delta_s)\sum_{\langle 2i-1,2i\rangle\parallel c}
      {\vec S}_{2i-1}\cdot{\vec S}_{2i}                     \nonumber \\
   &+&{\cal J}_{ab}^C\sum_{\langle ij\rangle\parallel ab} 
      {\vec S}_i\cdot{\vec S}_j.
\end{eqnarray}
Following the LSW theory, the spin wave dispersion is given by 
\begin{eqnarray}
\label{swc}
\omega_{C\pm}({\bf k})\!&=&\!
2\Big\{\big[2{\cal J}_{ab}+|{\cal J}_c|                   \nonumber \\
&\pm&\! {\cal J}_c(\cos^2k_z+\delta_s^2\sin^2k_z)^{1/2}\big]^2\!
-\big(2{\cal J}_{ab}\gamma_{\bf k}\big)^2\Big\}^{1/2}.    \nonumber \\
\end{eqnarray}
For the numerical evaluation of Fig. \ref{fig:swcafd}
we have used the experimental exchange interactions:\cite{Ulr03} 
${\cal J}_{ab}=2.6$ meV, ${\cal J}_c=3.1$ meV, $\delta_s=0.35$.
Indeed, large gap is found between two modes halfway in between the 
$M$ and $R$ points, and between the $Z$ and $\Gamma$ points (not shown),
respectively. Two modes measured\cite{Ulr03} and obtained from the 
present theory in the unfolded Brillouin zone follow from the dimerized 
magnetic structure.

The microscopic reason of the anisotropy in the exchange constants 
${\cal J}_{c1}$ and ${\cal J}_{c2}$ is the tendency of the orbital 
chain to dimerize, as we have demonstrated in Sec. \ref{sec:dim}. 
Such a dimerized orbital chain may only be stable, however, if the 
corresponding interactions in the orbital sector (\ref{jopm}) alternate, 
i.e., $\delta_o>0$. This becomes possible at finite temperature when 
also intersite spin correlations may alternate along the $c$ axis,
supporting such a dimerized state. Although a completely satisfactory 
treatment of the spin correlations in a broad temperature regime which 
covers the symmetry broaken $C$-AF phase is not possible at the moment,
we have employed the LSW theory to calculate the spin correlations 
$\langle{\vec S}_i\cdot{\vec S}_{i+1}\rangle$, as explained in the 
Appendix \ref{sec:quantum}. 

%%%%%%%%%%%%%%%%%%%%%%%%%%%%%%%%%%%%%%%%%%%%%%%%%%%%%%%%%%%%%%%%%%%%%%%%
%%                             Figure 13
%%%%%%%%%%%%%%%%%%%%%%%%%%%%%%%%%%%%%%%%%%%%%%%%%%%%%%%%%%%%%%%%%%%%%%%%
\begin{figure}
\includegraphics[width=7.5cm]{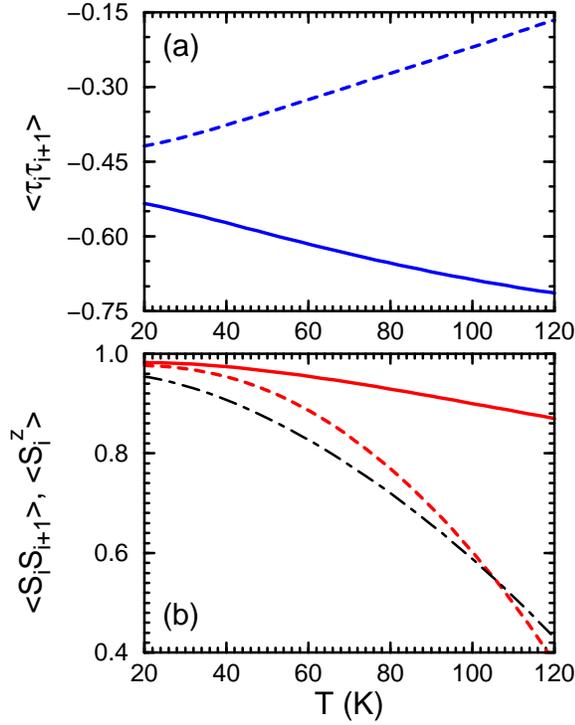}
\caption{(Color online)
Result of the self-consistent calculation of intersite correlations
in the dimerized $C$-AF phase along $c$ axis for increasing temperature:
(a) orbital $\langle{\vec\tau}_i\cdot{\vec\tau}_{i+1}\rangle$, and 
(b) spin    $\langle{\vec   S}_i\cdot{\vec   S}_{i+1}\rangle$. 
The correlations on stronger (weaker) FM bonds are shown by solid 
(dashed) lines. Dashed-dotted line in (b) shows the order parameter
$\langle S^z\rangle$ in the $C$-AF phase as obtained from the LSW 
theory. Parameters: $\eta=0.13$, $J=30$ meV, and $p=0.30$, 
see Fig. \ref{fig:jplaq}.
} 
\label{fig:dios}
\end{figure}

The result of the self-consistent calculation of spin and orbital 
correlations along the $c$ axis in the dimerized $C$-AF phase is shown 
in Fig. \ref{fig:dios}. The driving force to stabilize the dimerized 
state is the instability of the orbital chain which leads to rather 
strong anisotropy in the orbital correlations [Fig. \ref{fig:dios}(a)]. 
On the contrary, the spin correlations differ only by a rather small 
amount (unlike in the OVB phase), as the large spins $S=1$ are far less 
susceptible to follow the dimerized structure, and the long range spin 
order is supported by the exchange interactions in all three directions. 
The energy of the dimerized state is lower than that of the undimerized 
$C$-AF structure. Apparently, a weak anisotropy between 
$\langle{\vec S}_i\cdot{\vec S}_{i+1}\rangle\sim 0.93$ and $\sim 0.79$ 
on stronger/weaker FM bonds encountered at $T=77$ K [see Fig. 
\ref{fig:dios}(b)] is already sufficient to trigger a phase transition 
to this phase from the $G$-AF phase stable below $T_{N2}$. Why this 
transition may really happen in YVO$_3$ we explain in the following 
Section. 

%%%%%%%%%%%%%%%%%%%%%%%%%%%%%%%%%%%%%%%%%%%%%%%%%%%%%%%%%%%%%%%%%%%%%%%%
%%                            Mechanism 
%%%%%%%%%%%%%%%%%%%%%%%%%%%%%%%%%%%%%%%%%%%%%%%%%%%%%%%%%%%%%%%%%%%%%%%%
\subsection{ Mechanism of the phase transition from $G$-AF 
                                                 to $C$-AF phase }
\label{sec:trans}

The transition from $G$-AF to $C$-AF phase in YVO$_3$ is puzzling as the 
magnetic order changes completely at finite temperature $T_{N2}\simeq77$ 
K, and the magnetic moments reorient.\cite{Ren00} The observed change of 
the spin and orbital pattern indicates that the spin-orbital superexchange 
interactions are frustrated and it is easy to tip the balance of these 
interactions and to change completely both the magnetic and orbital order. 
As the transition between the two phases occurs at finite 
temperature, the entropy has to play an important role, so it was 
suggested before that the large orbital entropy due to orbital 
fluctuations in the $C$-AF phase could be released at $T_{N2}$ and
trigger the transition.\cite{Kha01} A closer inspection of the present
model and the reconsideration of recent experiments show, however, that 
the situation is somewhat more intricate. 

First of all, we have already emphasized that the magnetic exchange 
constants are reduced in the $C$-AF phase, and we presented a possible
mechanism responsible for this reduction in Sec. \ref{sec:red}. 
As a result of orbital fluctuations, the average energy of magnetic 
excitations is lowered in the $C$-AF phase (Fig. \ref{fig:cgaf}), so
one expects that the {\it spin entropy\/} might play an important role
as well. 
Using the spin and orbital excitations derived already for both phases
in the previous Sections, we estimate these entropy contributions 
assuming that the excitations are independent from each other. 
The spin waves are given by Eqs. (\ref{swg}) and (\ref{swc}), 
while the orbital excitations by Eqs. (\ref{owg}) and (\ref{owc0}).
Here we will ignore the change of the orbital excitations in the 
dimerized $C$-AF phase as this gives only a marginal contribution 
to the entropy of the $C$-AF phase, and does not influence the 
magnetic transition at $T_{N2}$ significantly. 

%%%%%%%%%%%%%%%%%%%%%%%%%%%%%%%%%%%%%%%%%%%%%%%%%%%%%%%%%%%%%%%%%%%%%%%%
%%                             Figure 14
%%%%%%%%%%%%%%%%%%%%%%%%%%%%%%%%%%%%%%%%%%%%%%%%%%%%%%%%%%%%%%%%%%%%%%%%
\begin{figure}
\includegraphics[width=7cm]{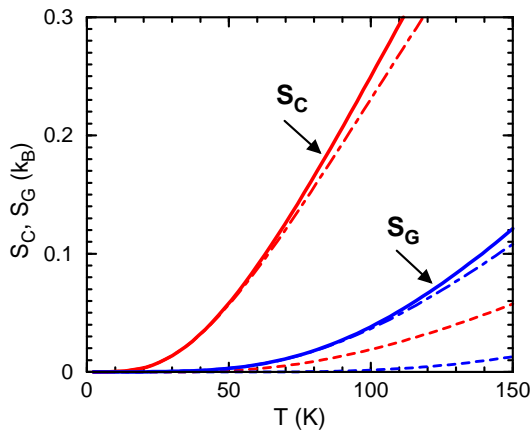}
\caption{(Color online)
Entropy of the $C$-AF ($S_C$) and $G$-AF ($S_G$) phase as obtained for 
the spin-orbital model (\ref{model}) using the experimental values of 
magnetic exchange constants in both phases. 
The dominating contributions result from spin excitations (dashed-dotted
lines), while the orbital contributions (dashed lines) are much smaller,
but give also a higher entropy in the $C$-AF phase. 
Parameters: $J=40$ meV, $\eta=0.13$, $V_a=0.30J$, $V_c=0.84J$.  
} 
\label{fig:ent}
\end{figure}

The spin and orbital entropy normalized per one vanadium ion is 
calculated using standard formulae:  
\begin{eqnarray}
\label{SC}
{\cal S}_C&=&k_BT\frac{1}{2N}\sum_{\bf k}\Big\{
   \log\Big(1-e^{-\beta\omega_{C+}({\bf k})}\Big)         \nonumber \\
&+&\log\Big(1-e^{-\beta\omega_{C-}({\bf k})}\Big)\Big\}   \nonumber \\
&+&k_BT\frac{1}{N_1}\sum_k \log\Big(1-e^{-\beta\Omega_{C}(k)}\Big), \\
\label{SG}
{\cal S}_G&=&k_BT\frac{1}{N}\sum_{\bf k}\{
   \log\Big(1-e^{-\beta\omega_{G}({\bf k})}\Big)          \nonumber \\
&+&k_BT\frac{1}{N_1}\sum_k \log\Big(1-e^{-\beta\Omega_{G}(k)}\Big),          
\end{eqnarray}
where $\beta=1/k_BT$, and $N$ ($N_1$) is the number of ${\bf k}$ ($k$) 
values. The entropy consists of the spin and orbital entropy terms for 
each phase. All summations are over the Brillouin zone which 
corresponds to the undimerized $C$-AF phase. Using the parameters 
consistent with the experimental data of Ulrich {\it et al.\/}
\cite{Ulr03} one finds (see Fig. \ref{fig:ent}) that: 
 (i) the entropy ${\cal S}_C$ for the $C$-AF phase is larger that 
     ${\cal S}_G$ for the $G$-AF phase, and 
(ii) the spin entropy grows significantly faster with temperature than 
     the orbital entropy for each phase.      
Therefore, we conclude that the spin entropy gives here a more 
important contribution and decreases the difference between the free 
energies of both magnetic phases in the temperature range $T\sim T_{N2}$.

%%%%%%%%%%%%%%%%%%%%%%%%%%%%%%%%%%%%%%%%%%%%%%%%%%%%%%%%%%%%%%%%%%%%%%%%
%%                             Figure 15
%%%%%%%%%%%%%%%%%%%%%%%%%%%%%%%%%%%%%%%%%%%%%%%%%%%%%%%%%%%%%%%%%%%%%%%%
\begin{figure}[t!]
\includegraphics[width=7cm]{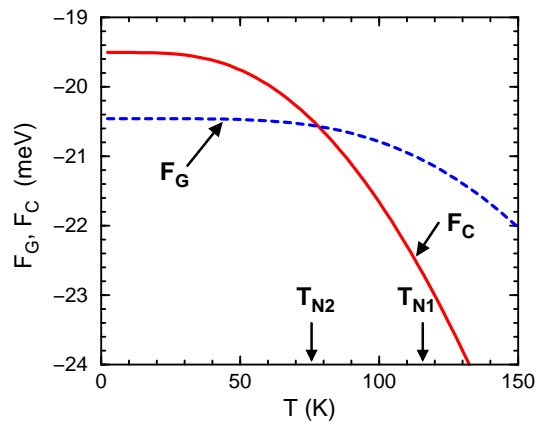}
\caption{(Color online)
Free energies of the $C$-AF (${\cal F}_C$, solid line) and $G$-AF 
(${\cal F}_G$, dashed line) phase as obtained for the spin-orbital model 
(\ref{model}) using the experimental values of magnetic exchange 
constants\cite{Ulr03} in both phases. The experimental magnetic 
transition temperatures, 
$T_{N2}\simeq 77$ K and 
$T_{N1}\simeq 116$ K, are indicated by arrows. 
Parameters are the same as in Fig. \ref{fig:ent}.  
} 
\label{fig:fene}
\end{figure}

It has been argued before\cite{Kha01,Kha05} that the difference between 
the energies of both phases, $E_G$ and $E_C$, has to be small at $T=0$. 
Indeed, we evaluated the free energy of both phases using the above 
entropies (\ref{SC}) and (\ref{SG}),
\begin{eqnarray}
\label{FC}
{\cal F}_C&=&E_C-T{\cal S}_C,               \\
\label{FG}
{\cal F}_G&=&E_G-T{\cal S}_G,              
\end{eqnarray}
and found that $E_C-E_G\simeq 1$ meV, and the transition from $G$-AF to 
$C$-AF phase is reproduced at the experimental value of the temperature 
$T_{N2}$ when the orbital interactions are chosen properly. 
In Fig. \ref{fig:fene} we show a representative case with $V_a=0.30J$, 
$V_c=0.84J$ with $J=40$ meV. Of course, this fit is not unique and 
$V_a$ ($V_c$) could be somewhat smaller (larger), but the energy 
difference $E_C-E_G$ at $T=0$ remains close to 1 meV in all cases.
Note, however, that too large values of $V_c$ are not allowed, as then
the $C$-AF phase gets destabilized by orbital excitations (\ref{owc0}).

%%%%%%%%%%%%%%%%%%%%%%%%%%%%%%%%%%%%%%%%%%%%%%%%%%%%%%%%%%%%%%%%%%%%%%%%
%%                             Finito
%%%%%%%%%%%%%%%%%%%%%%%%%%%%%%%%%%%%%%%%%%%%%%%%%%%%%%%%%%%%%%%%%%%%%%%%
\section{ Summary and conclusions }
\label{sec:finale}

The present study shows the importance of $t_{2g}$ orbital degrees of 
freedom in cubic vanadates. We have presented the spin-orbital model 
for cubic vanadates and analyzed its possible solutions in various 
parameter regimes, using extensively the decoupling of spin and orbital 
degrees of freedom. Although the model is more general, we have focused 
on the solutions which arise in the case of anisotropic occupancy of 
$t_{2g}$ orbitals, with $xy$ orbitals singly occupied at each site. This state is 
believed to be realized in cubic vanadates, at least in low temperature 
phases with magnetic long range order. When Hund's exchange and orbital 
interactions promoted by the lattice are weak, the superexchange is 
strongly frustrated and gives a rather exotic dimerized OVB state, with 
orbital singlets alternating along the $c$ axis and stabilized at every 
second bond by ferromagnetic spin correlations. 
In this way, spin and orbital correlations support each other and 
demonstrate a unique instability of the spin-orbital system towards 
a dimerized state.\cite{Hor03,Sir03} This instability turns out to play 
also an important role at finite temperature in YVO$_3$, but in
a different regime of parameters where its mechanism is more subtle.

When Hund's exchange or the orbital interactions increase, the OVB 
ground state is disfavored and a particular type of long range magnetic 
order emerges instead from the frustrated superexchange interactions in 
cubic vanadates. These other AF states ($C$-AF and $G$-AF phases), 
as well as the OVB state itself (at low $J_H$), demonstrate a close 
interrelation between magnetic and orbital order, 
with complementary behavior of spin and 
orbital correlations, known as the so-called Goodenough-Kanamori rules.
\cite{Goo63,Kan59} While in some cases these rules (and the underlying 
decoupling of spin and orbital operators) work well, we have presented 
the case of the $C$-AF phase with rather disordered orbitals, where it 
is likely that joint spin-orbital fluctuations also play a role,
\cite{Ole06} and it would be necessary to include them for a more 
quantitative comparison with experiment. 

A detailed analysis of the possible solutions of the spin-orbital 
superexchange model supplemented by the orbital interactions induced by
the lattice demonstrates that two different types of AF order, $C$-AF 
and $G$-AF phase, compete with each other in the parameter regime 
relevant for YVO$_3$. However, the energetic proximity of these two 
phases in a particular parameter regime could explain possible changes
of magnetic order by pressure or magnetic field --- when the microscopic 
parameters are fixed, one or the other phase could be stable at 
$T\to 0$. The situation changes at finite temperature, however, when the 
{\it spin and orbital excitations\/} are of importance and may tip the 
energy balance between given two types of order by the entropy term. 
In fact, we have shown that this is likely to be the microscopic 
explanation of the observed first-order phase transition and 
switching of the magnetic order in YVO$_3$ at $T_{N2}$.    

Our study has established that the nature of the transition from the
$G$-AF to $C$-AF phase at $T_{N2}$ observed in YVO$_3$ is complex and 
several factors have to come together to trigger it when temperature 
increases: 
(i) the presence of active $t_{2g}$ orbital degrees of freedom opens a 
possibility of two different types of AF order which may
compete with each other;
(ii) rigidity of the $C$-AO order in the $G$-AF phase hampers possible
free energy gains when spin or orbital excitations are created (as spin 
interactions are rather strong and the orbital gap is quite large);
(iii) the change of structure observed at $T_{N2}$ not only helps to 
stabilize the weak $G$-AO order, but also releases more orbital 
fluctuations when the $xy$ {\it orbitals become active\/} and their 
occupancy is not fixed --- such fluctuations result in turn in 
fluctuating magnetic exchange constants and lead to the reduction of 
the characteristic energy scale for the spin excitations, and finally
(iv) the spin correlations have to be weakened by increasing 
temperature to participate in a joint spin-orbital dimerization in
the $C$-AF phase. 
Thus, the difference between the $G$-AF and $C$-AF phase of YVO$_3$ is 
much deeper than simply the observed difference in the magnetic order. 
It is far more important that the {\it orbital state softens\/} at the 
transition at $T_{N2}$ to the $C$-AF phase and this change happens in 
a concerted way with the observed reorientation of the magnetic moments. 
In addition, the intrinsic instability in the orbital sector towards 
dimerization, which is incompatible with the magnetic order in the $G$-AF 
phase and is blocked by spin correlations in the $C$-AF phase at $T=0$, 
becomes possible when the intersite ferromagnetic spin correlations 
along $c$ axis have been somewhat weakened with increasing temperature. 
  
Although we have suggested a plausible scenario of the observed 
magnetic phase transition in YVO$_3$, the microscopic theory of the 
$C$-AF phase at finite temperature remains still to be constructed. 
It is not clear at the moment to what extent the $xy$ orbital
fluctuations are released in the intermediate magnetic phase and are 
still present up to the structural transition at $T_s\sim 200$ K (see
Fig. 1). It could well be that spin-orbital {\it entanglement in excited
states\/} plays a role in this temperature regime and prevents reliable 
evaluation of the magnetic exchange constants in the $C$-AF phase using 
the conventional decoupling of spin and orbital operators. Furthermore, 
it is puzzling whether dimerization also plays a role in reducing the 
the magnetic order parameter in the $C$-AF phase which is hard to 
explain using the spin-wave theory, or the above entanglement is the
main reason responsible for this reduction. Note however that it could 
be argued that the observed orientation of the magnetic moments which are close to 
lying within the $ab$ planes is enforced by the dimerization in the 
$C$-AF phase.

Some other problems remain still open and should be treated in future 
more complete theory. We note that also in the $G$-AF phase a 
considerable reduction of the magnetic order parameter\cite{Ulr03} goes 
beyond that expected from the quantum fluctuations.\cite{Rac02} We believe that the 
relativistic spin-orbit coupling $\propto\lambda{\vec L}_i{\vec S}_i$ 
contributes significantly to the magnetic properties 
in the entire regime of temperature,\cite{Hor03} in particular also to the spin 
correlations in the $G$-AF phase, and could reduce the observed value of 
the magnetization. In fact, it would also break the symmetry in the spin 
space and determine an easy axis for the AF order parameter. 
In the present study the spin-orbit coupling $\lambda$ was ignored, as 
in the considered regime of $J\gg\lambda$ it could lead just to the 
perturbative corrections of the presented spin and orbital excitations. 
In contrast, in the regime of $\lambda\sim J$ it would lead to ordering 
of orbitals with complex coefficients, 
$(|xz\rangle\pm i|yz\rangle)/\sqrt{2}$, 
with finite orbital angular momentum.\cite{Hor03} Although the cubic
vanadates are not in this regime of parameters, finite spin-orbit 
coupling $\lambda$ would be crucial for quantitative understanding of 
spin (and orbital) excitations in the entire parameter regime. This
interaction provides another mechanism for the softening of spin
excitations in the $C$-AF phase, which would complement the scenario
considered in this paper. We also note that a small $G$-like 
magnetization component was observed as well in the $C$-AF phase in the 
temperature range $T_{N2}<T<T_{N1}$. Therefore, it is likely that this magnetic 
phase is still more complex than suggested in the present paper, and 
requires a more careful analysis. Recent progress in experimental 
methods makes it possible to measure also orbital excitations,
\cite{Ish00,Ulr06} and information on the orbital excitations in 
YVO$_3$ would be instrumental to resolve some of the above problems. 

Summarizing, we have presented the consequences of the microscopic 
spin-orbital model in the parameter regime relevant for cubic vanadates
and suggested a scenario which explains the magnetic transition between 
the $G$-AF and dimerized $C$-AF phases observed in YVO$_3$. This study 
indicates a close relationship between the observed magnetic 
correlations in the ground state and the structural transition, which 
in case of YVO$_3$ occurs well above the first magnetic transition. 
Thus we conclude that a careful analysis of the mechanism of the 
structural transition and its dependence on the actual chemical
composition is challenging and needed for complete theoretical 
understanding of the experimental phase diagram of cubic vanadates.

%%%%%%%%%%%%%%%%%%%%%%%%%%%%%%%%%%%%%%%%%%%%%%%%%%%%%%%%%%%%%%%%%%%%%%%%
%%
%%                            ACKNOWLEDGMENTS
%%
%%%%%%%%%%%%%%%%%%%%%%%%%%%%%%%%%%%%%%%%%%%%%%%%%%%%%%%%%%%%%%%%%%%%%%%%
\acknowledgments

We thank B. Keimer and C. Ulrich for stimulating and insightful 
discussions. 
A.~M.~Ole\'s would like to acknowledge support by the Polish Ministry
of Science and Education under Project No.~N202 068 32/1481.

%%%%%%%%%%%%%%%%%%%%%%%%%%%%%%%%%%%%%%%%%%%%%%%%%%%%%%%%%%%%%%%%%%%%%%%%
%%
%%                             APPENDICES
%%
%%%%%%%%%%%%%%%%%%%%%%%%%%%%%%%%%%%%%%%%%%%%%%%%%%%%%%%%%%%%%%%%%%%%%%%%
\appendix

%%%%%%%%%%%%%%%%%%%%%%%%%%%%%%%%%%%%%%%%%%%%%%%%%%%%%%%%%%%%%%%%%%%%%%%%
%%                           Derivation
%%%%%%%%%%%%%%%%%%%%%%%%%%%%%%%%%%%%%%%%%%%%%%%%%%%%%%%%%%%%%%%%%%%%%%%%
\section{ Derivation of the spin-orbital model }
\label{sec:derivation}

The effective superexchange interactions between two V$^{3+}$ ions in 
$d^2$ configuration with spin $S=1$ (triplet $^3T_2$ state) at sites $i$ 
and $j$ for a bond $\langle ij\rangle$ oriented along one of the cubic 
axes $\gamma=a,b,c$ originate from virtual charge excitations by the 
hopping processes which involve two active $t_{2g}$ orbitals along its 
direction. As an example, we consider here a bond along $c$ axis 
($\gamma=c$), with active $a$ ($yz$) and $b$ ($xz$) orbitals. In this 
case the charge excitation by either $a$ or $b$ electron leads to one of
three possible $d^3$ 
excited states: $abc$, $a^2c$ or $b^2c$ (see Fig. \ref{fig:se}).
The actual configuration $c^1$ of the inactive orbital $c$ enters via the 
constraint (\ref{const}). The total spin per two sites is conserved in 
the $d^2_id^2_j\rightarrow d^3_id^1_j$ excitation process, i.e., the 
electron transferred in the excitation process and two other 
electrons on the $d^3$ site are either in high-spin 
($S=\frac{3}{2}$) state, or in low-spin ($S=\frac{1}{2}$) state. 
Therefore, when the second order processes $d^2_id^2_j\rightarrow d^3_i
d^1_j\rightarrow d^2_id^2_j$ are analyzed, one has to project the $d^3_i$ 
configuration generated after an individual hopping process on the 
respective $d^3_i$ eigenstates. Similar, when a deexcitation process took 
place, one has to project the resulting $d^2_i$ configuration on the 
triplet $^3T_2$ ground state. 

%%%%%%%%%%%%%%%%%%%%%%%%%%%%%%%%%%%%%%%%%%%%%%%%%%%%%%%%%%%%%%%%%%%%%%%%
%%                         general structure
%%%%%%%%%%%%%%%%%%%%%%%%%%%%%%%%%%%%%%%%%%%%%%%%%%%%%%%%%%%%%%%%%%%%%%%%
The general form of the effective Hamiltonian follows from symmetry 
considerations for the possible $d^2_id^2_j\rightarrow d^3_id^1_j
\rightarrow d^2_id^2_j$ processes which contribute to the superexchange.
The total spin states in the excited states are well described by the
spin operators:
\begin{eqnarray}
\label{spinpro}
{\cal P}_{\rm HS}({\vec S}_i,{\vec S}_j)=
         \frac{1}{S(2S+1)}({\vec S}_i\cdot {\vec S}_j+2),     \\
{\cal P}_{\rm LS}({\vec S}_i,{\vec S}_j)=
         \frac{1}{S(2S+1)}({\vec S}_i\cdot {\vec S}_j-1),
\end{eqnarray}
which correspond to the high-spin 
${\cal P}_{\rm HS}({\vec S}_i,{\vec S}_j)$ and low-spin 
${\cal P}_{\rm LS}({\vec S}_i,{\vec S}_j)$ excited states, respectively. 
The orbital state is described by the orbital operators 
${\cal Q}_n(i,j)$, where $n$ refers to different excited states.
\cite{Kha04} In the present case $n=1$ corresponds to high-spin 
$S=\frac{3}{2}$ excited states in Fig. \ref{fig:se}(a), $n=2$ to 
low-spin $S=\frac{1}{2}$ excited states in Fig. \ref{fig:se}(b), while 
$n=3$ describes the orbital state realized for the excitations of double 
occupancies shown in Fig. \ref{fig:se}(c). As the orbital quantum number 
is conserved along the hopping process (\ref{hkin}), either the same two 
orbitals are occupied before and after the virtual excitation, or an 
orbital fluctuation shown in Fig. \ref{fig:se} takes place, and the 
occupied orbitals are interchanged between sites $i$ and $j$. The latter 
processes are unique for the $t_{2g}$ orbitals and do not occur for 
degenerate and singly occupied $e_g$ orbitals, where the orbital quantum 
number is not conserved and single orbital excitations are possible 
instead.\cite{Ole00}

%%%%%%%%%%%%%%%%%%%%%%%%%%%%%%%%%%%%%%%%%%%%%%%%%%%%%%%%%%%%%%%%%%%%%%%%
%%                        interactions along c
%%%%%%%%%%%%%%%%%%%%%%%%%%%%%%%%%%%%%%%%%%%%%%%%%%%%%%%%%%%%%%%%%%%%%%%%
In the case of cubic vanadates one arrives therefore at a general 
expression, 
\begin{eqnarray}
\label{allv}\hskip -.4cm
{\cal H}&=&\!\!\sum_{\langle ij\rangle}\Big\{
 - \frac{1}{3}\frac{t^2}{\varepsilon(^4A_2)}
   ({\vec S}_i\cdot {\vec S}_j+2){\cal Q}_1(i,j)  \nonumber \\
 &+&\!\! \frac{1}{3}\frac{t^2}{\varepsilon(^2E)}
   ({\vec S}_i\cdot {\vec S}_j-1){\cal Q}_2(i,j)  \nonumber \\
 &+&\!\! \frac{1}{2}\left(\frac{t^2}{\varepsilon(^2T_1)}
                                +\frac{t^2}{\varepsilon(^2T_2)}\right)
   ({\vec S}_i\cdot {\vec S}_j-1){\cal Q}_3(i,j)\Big\},
\end{eqnarray}
The first term $\propto{t^2}/{\varepsilon(^4A_2)}$ is FM, while the 
remaining terms stand for different AF contributions.
The coefficient 1/2 in the contributions due to $^2T_1$ and $^2T_2$
excited states follows from the projection of the double occupancies 
of one of the active orbitals, either $a_i^2c_i$ or $b_i^2c_i$ 
[Fig. \ref{fig:se}(c)], onto the eigenstates of V$^{2+}$ ions. 
The orbital states which contribute 
to the above structure of superexchange (\ref{allv}) depend on the bond
direction; here we give as an example a complete expression for the 
bonds $\langle ij\rangle$ along the $c$-direction,
\begin{widetext}
\begin{eqnarray}
\label{sexc}
{\cal H}_c&=&\sum_{\langle ij\rangle\parallel c}\Big\{\;
-\frac{1}{3}\;\frac{2t^2}{\varepsilon(^4A_2)}
({\vec S}_i\cdot {\vec S}_j+2)\;[(1-n_{ia})(1-n_{jb})
+(1-n_{ib})(1-n_{ja})-(a_i^{\dagger}b_i^{}b_j^{\dagger}a_j^{}
+b_i^{\dagger}a_i^{}a_j^{\dagger}b_j^{})n_{ic}n_{jc}]        \nonumber \\
&-&\frac{1}{3}\;\frac{t^2}{\varepsilon(^4A_2)}\;
({\vec S}_i\cdot {\vec S}_j+2)\;
\left[(1-n_{ic})n_{jc}+n_{ic}(1-n_{jc})\right]               \nonumber \\  
&+&\frac{1}{3}\;\frac{2t^2}{\varepsilon(^2E)}\;
({\vec S}_i\cdot {\vec S}_j-1)\;
[(1-n_{ia})(1-n_{jb})+(1-n_{ib})(1-n_{ja})
+\frac{1}{2}(a_i^{\dagger}b_i^{}b_j^{\dagger}a_j^{}
            +b_i^{\dagger}a_i^{}a_j^{\dagger}b_j^{})n_{ic}n_{jc}] 
                                                             \nonumber \\
&+&\frac{1}{3}\;\frac{t^2}{\varepsilon(^2E)}\;
({\vec S}_i\cdot {\vec S}_j-1)\;
\left[(1-n_{ic})n_{jc}+n_{ic}(1-n_{jc})\right]               \nonumber \\  
&+&\frac{1}{2}\;\left(\frac{t^2}{\varepsilon(^2T_1)}+
                      \frac{t^2}{\varepsilon(^2T_2)}\right)\;
		      ({\vec S}_i\cdot {\vec S}_j-1)\;
\left[(1-n_{ia})(1-n_{jb})+(1-n_{ib})(1-n_{ja})-n_{jc}-n_{jc}+2\right]
                                                             \nonumber \\ 
&+&\frac{1}{2}\;\left(\frac{t^2}{\varepsilon(^2T_1)}-
                      \frac{t^2}{\varepsilon(^2T_2)}\right)\;
		      ({\vec S}_i\cdot {\vec S}_j-1)\;
\left(a_i^{\dagger}b_i^{}a_j^{\dagger}b_j^{}+
      b_i^{\dagger}a_i^{}b_j^{\dagger}a_j^{}\right)\Big\}.  
\end{eqnarray}
\end{widetext}
The operators $a_i^{\dagger}$ and $b_i^{\dagger}$ are (spinless) fermion
creation operators in the active orbitals $|a\rangle$ and $|b\rangle$ at
site $i$, respectively, while $n_{i\gamma}^{}=\gamma_i^{\dagger}\gamma_i^{}$ 
are fermion number operators in state $|\gamma\rangle$ at site $i$, with
$\gamma=a,b,c$.  

%%%%%%%%%%%%%%%%%%%%%%%%%%%%%%%%%%%%%%%%%%%%%%%%%%%%%%%%%%%%%%%%%%%%%%%%
%%                     interactions along a and b
%%%%%%%%%%%%%%%%%%%%%%%%%%%%%%%%%%%%%%%%%%%%%%%%%%%%%%%%%%%%%%%%%%%%%%%%
The effective interactions on the bonds within the $ab$ planes may 
be now obtained by rotating Eq. (\ref{sexc}) to the bonds oriented 
along either $a$ or $b$ axis. Note that the orbital operators which
correspond to the active $|a\rangle$ and $|b\rangle$ orbitals are then
replaced by either $|b\rangle$ and $|c\rangle$ (for $a$ axis), or by
$|a\rangle$ and $|c\rangle$ (for $b$ axis), while the general structure
of the superexchange Hamiltonian (\ref{sexc}) remains the same.

%%%%%%%%%%%%%%%%%%%%%%%%%%%%%%%%%%%%%%%%%%%%%%%%%%%%%%%%%%%%%%%%%%%%%%%%
%%                        interactions at J_H\to 0
%%%%%%%%%%%%%%%%%%%%%%%%%%%%%%%%%%%%%%%%%%%%%%%%%%%%%%%%%%%%%%%%%%%%%%%%
As both FM and AF terms are present in Eq. (\ref{sexc}), the 
superexchange interactions are frustrated. It is instructive to consider 
the limit of $J_H\to 0$ in which the multiplet structure of V$^{2+}$ 
ions collapses to a single excitation energy $U$. In this case the 
interactions simplify considerably, and the terms $\propto n_{i\mu}$ 
which originate from the excitations with three different orbitals 
occupied at the same site cancel each other. 
This feature is analogous to the similar compensation of the high- and 
low-spin processes in the superexchange models with degenerate $e_g$ 
orbitals.\cite{Ole01} However, as a new feature one finds a nonvanishing
contribution due to the orbital fluctuations, 
$\propto (a_i^{\dagger}b_i^{}b_j^{\dagger}a_j^{}+H.c.)$, as the terms
which originate from the high- and low-spin processes add to each other.
As usual, the double occupancies in the excited states lead to the AF 
terms as a consequence of the Pauli principle. These simplifications 
lead to the following form of the effective Hamiltonian along the $c$ 
axis in the limit of $J_H\to 0$,
\begin{eqnarray}
\label{sex0}
{\cal H}_c(\eta=0)&=&J\sum_{\langle ij\rangle\parallel c}
({\vec S}_i\cdot {\vec S}_j+1)                          \nonumber \\
&\times&\Big(n_{ia}n_{jb}+n_{ib}n_{ja}    
  + a_i^{\dagger}b_i^{}b_j^{\dagger}a_j^{}
  + b_i^{\dagger}a_i^{}a_j^{\dagger}b_j^{}\Big),        \nonumber \\
\end{eqnarray}
where $J=4t^2/U$ is the superexchange interaction. This expression may 
be also written in a more compact form,
\begin{equation}
\label{sextau}
{\cal H}_c(\eta=0)=\frac{1}{2}J\sum_{\langle ij\rangle\parallel c}
 ({\vec S}_i\cdot   {\vec S}_j+1)\Big(
{\vec\tau}_i\cdot {\vec\tau}_j+\frac{1}{4}n_in_j\Big),
\end{equation}
where the scalar product of orbital pseudospin operators 
${\vec\tau}_i=\{\tau_i^+,\tau_i^-,\tau_i^z\}$ is defined by:
\begin{eqnarray}
\label{tauz}
\tau_i^+&\equiv &a_i^{\dagger}b_i^{},\hskip .7cm 
\tau_i^- \equiv  b_i^{\dagger}a_i^{},
\nonumber \\
\tau_i^z&\equiv &\frac{1}{2}(n_{ia}-n_{ib}).
\end{eqnarray}
Here we use spinless fermion operators $a_i^{\dagger}$ and 
$b_i^{\dagger}$, but one could also introduce instead Schwinger boson
operators.
For the bonds along either $a$ or $b$ axis similar expressions 
obtained from Eqs. (\ref{tauz}) by cyclic permutations of the orbitals 
$\{a,b,c\}$ have to be used. If in addition the $c$ orbitals are 
condensed ($n_{ic}=1$), as in YVO$_3$, one finds a simplified form of 
Eq. (\ref{sextau}) for the bonds along $c$ axis,
\begin{equation}
\label{sexc1}
{\cal H}_{c}^{(0)}=\frac{1}{2}J\sum_{\langle ij\rangle\parallel c}
 ({\vec S}_i\cdot   {\vec S}_j+1)\Big(
{\vec\tau}_i\cdot {\vec\tau}_j+\frac{1}{4}\Big).
\end{equation}
The orbital interactions are then purely classical on the bonds in $ab$ 
plane as 
$({\vec\tau}_i\cdot{\vec\tau}_j+\frac{1}{4}n_in_j)^{ab}\equiv\frac{1}{2}$ 
for these bonds.
 
%%%%%%%%%%%%%%%%%%%%%%%%%%%%%%%%%%%%%%%%%%%%%%%%%%%%%%%%%%%%%%%%%%%%%%%%
%%                         Quantum corrections 
%%%%%%%%%%%%%%%%%%%%%%%%%%%%%%%%%%%%%%%%%%%%%%%%%%%%%%%%%%%%%%%%%%%%%%%%
\section{ Spin and orbital excitations  
          in the dimerized $C$-AF phase }
\label{sec:quantum}

Here we explain the full algebraic structure of the spin and orbital 
wave problem in the dimerized $C$-AF phase. Its solution gives the 
both types of excitation energies, and provides a systematic method to 
evaluate both the value of the order parameter,
\begin{equation}
\langle S^z_i\rangle\equiv S-\delta S^z,
\label{szco}
\end{equation}
and the intersite spin correlations along $c$ axis, 
\begin{equation}
\langle{\vec S}_i\cdot{\vec S}_{i+1}\rangle\equiv {\cal C}_{i,i+1}=
{\cal C}_0+\delta{\cal C}e^{i\pi z_i},
\label{ssco}
\end{equation}
where $z_i$ is the $z$th coordinate of the vactor $R_i$ corresponding 
to site $i$. If exchange interactions alternate along $c$ axis, as 
given by Eqs. (\ref{Jcp1}) and (\ref{Jcp2}), the alternating part 
$\delta{\cal C}$ of the intersite spin correlation function is finite.  

In order to evaluate the order parametr (\ref{szco}) and the
intersite spin correlations (\ref{ssco}) in the dimerized $C$-AF phase
within the LSW formalism it is convenient to arrive first at the 
boson representation of the spin Hamiltonian. Therefore, we performed 
the transformation to a ferromagnet (\ref{rota}) and the subsequent 
Holstein-Primakoff transformation (\ref{bosons}) to 
$\{a_i,a_i^{\dagger}\}$ bosons. One finds the following form of the 
above averages,
\begin{eqnarray}
\label{szcoa}
\delta S^z&=&S-\langle a^{\dagger}_ia^{}_i\rangle, \\
{\cal C}_{i,i+1}&=&
\label{sscoa}
S^2-2\delta S^z+\langle a^{\dagger}_ia^{\dagger}_{i+1}\rangle,
\end{eqnarray}
and for the quadratic (LSW) Hamiltonian 
\begin{eqnarray}
\label{hcaflsw}
H_{\rm LSW}&=&
      {\cal J}_{ab}^C\sum_{\langle ij\rangle\parallel ab} 
      (n_i+n_j+a_i^{\dagger}a_j^{\dagger}+a_i^{}a_j^{})  \nonumber \\
&+&{\cal J}_{c}^C\sum_{\langle i,i+1\rangle\parallel c}  
\Big(1+e^{i\pi z_i}\delta_s\Big)                         \nonumber \\
&\times&\big(n_{i}+n_{i+1}
   -a_{i+1}^{\dagger}a_{i}^{}-a_{i}^{\dagger}a_{i+1}^{}),          
\end{eqnarray}
where ${\cal J}_{c}^C$ is the average value (\ref{Jcp}) of the FM 
exchange interaction along $c$ axis. 

Next we employ the Fourier transformation to boson operators in
reciprocal (momentum) space
\begin{equation}
a_{\bf k}^{\dagger}=\frac{1}{\sqrt{N}}\sum_i 
                    e^{ i{\bf k}{\bf R}_i}a_i^{\dagger},   \hskip .3cm
a_{\bf k}^{}       =\frac{1}{\sqrt{N}}\sum_i 
                    e^{-i{\bf k}{\bf R}_i}a_i^{},
\label{deco}
\end{equation}
which gives the LSW Hamiltonian in reciprocal space,
\begin{eqnarray}
\label{hcaflswk}
H_{\rm LSW}\!&=&\sum_{\bf k}\Big\{
4{\cal J}_{ab}^Ca_{\bf k}^{\dagger}a_{\bf k}^{}
 + 2{\cal J}_{ab}^C\gamma({\bf k})
                  (a_{\bf k}^{\dagger}a_{-{\bf k}}^{\dagger}
                   +a_{-{\bf k}}^{}   a_{\bf k}^{})  \nonumber \\
&+&2{\cal J}_{c}^C\big[(1-\cos k_z)a_{\bf k}^{\dagger}a_{\bf k}^{}
    +i\delta_s\sin k_z a_{\bf k}^{\dagger}a_{{\bf k}+{\bf Q}}^{}\big]\Big\},
                                                           \nonumber \\
\end{eqnarray}
where ${\bf Q}=(0,0,\pi)$ is the wave vector which corresponds to
the doubling of the unit cell along $c$ axis due to the dimerized 
$C$-AF spin structure. 

In order to find both the energies of spin wave excitations and the
average values of the correlation functions at finite temperature $T$, 
we introduce here temperature Green's functions for boson operators in 
the momentum space using the notation of Zubarev.\cite{Zub60,Hal72} 
The first of them satisfies the following equation of motion,
\begin{equation}
\omega\langle\langle a_{\bf k}^{}|a_{\bf k}^{\dagger}\rangle\rangle_{\omega} =
\frac{1}{2\pi}+\langle\langle 
 [a_{\bf k}^{},H_{\rm LSW}]|a_{\bf k}^{\dagger}\rangle\rangle_{\omega}.
\label{eqmo}
\end{equation}
It depends on energy $\omega$ and generates three more Green functions:
$\langle\langle a_{ {\bf k}+{\bf Q}}^{}|a_{\bf k}^{\dagger}\rangle\rangle_{\omega}$,
$\langle\langle a_{-{\bf k}        }^{}|a_{\bf k}^{\dagger}\rangle\rangle_{\omega}$,
and
$\langle\langle a_{-{\bf k}+{\bf Q}}^{}|a_{\bf k}^{\dagger}\rangle\rangle_{\omega}$.
In is next convenient to introduce the following expressions which 
define the algebraic structure of the spin wave problem:
\begin{eqnarray}
\label{abd}
  A_{{\bf k}\pm}&=&2{\cal J}_{c}^C(1\pm\cos k_z)+4{\cal J}_{ab}^C, \\
     B_{{\bf k}}&=&4{\cal J}_{ab}^C\gamma({\bf k}),                \\
\Delta_{{\bf k}}&=&2{\cal J}_{c}^C\delta_s\sin k_z.
\end{eqnarray}
The respective system of equations of motion generated by Eq. 
(\ref{eqmo}) is:
\begin{widetext}
\begin{equation}
\label{gfeq}
\left(\begin{array}{cccc}
 A_{{\bf k}-}-\omega_C({\bf k}) & i\Delta_{\bf k} & B_{\bf k} & 0    \\
-i\Delta_{\bf k} & A_{{\bf k}+}-\omega_C({\bf k}) & 0 & B_{\bf k}    \\
-B_{\bf k} & 0 & -A_{{\bf k}-}-\omega_C({\bf k})  & -i\Delta_{\bf k} \\
 0 & -B_{\bf k} & i\Delta_{\bf k} & -A_{{\bf k}+}-\omega_C({\bf k})
\end{array} \right)\left( 
\begin{array}{c}
\langle\langle a_{ {\bf k}}          |a_{\bf k}^{\dagger}\rangle\rangle_{\omega} \\
\langle\langle a_{ {\bf k}+{\bf Q}}  |a_{\bf k}^{\dagger}\rangle\rangle_{\omega} \\
\langle\langle a_{-{\bf k}}^{\dagger}|a_{\bf k}^{\dagger}\rangle\rangle_{\omega} \\
\langle\langle a_{-{\bf k}+{\bf Q}}^{\dagger}
                                     |a_{\bf k}^{\dagger}\rangle\rangle_{\omega}
\end{array} 
\right) = -\frac{1}{2\pi}\left( 
\begin{array}{c}
 1   \\
 0   \\
 0   \\
 0
\end{array} \right).
\end{equation}
%\end{widetext}
%
Eq. (\ref{gfeq}) has a typical structure obtained for elementary 
excitations in the random phase approximation (or in the LSW theory) 
for an antiferromagnet. One finds two positive eigenvalues 
$\omega_{C\pm}({\bf k})$ given by Eq. (\ref{swc}), and two negative 
ones, $-\omega_{C\pm}({\bf k})$. 

By solving the system of Eqs. (\ref{eqmo}) one finds the following 
Green's functions:
%
%\begin{widetext}
\begin{eqnarray}
\label{gf1}
\langle\langle a_{ {\bf k}}          |a_{\bf k}^{\dagger}\rangle\rangle_{\omega}&=&
+\frac{1}{2\pi}
 \frac{(\omega^2-A_{{\bf k}+}^2+B_{\bf k}^2)(\omega+A_{{\bf k}-})
      -\Delta_{\bf k}^2(\omega-A_{{\bf k}+})}
{\{\omega^2-\omega_{C+}^2({\bf k})\}\{\omega^2-\omega_{C-}^2({\bf k})\}}, \\
\langle\langle a_{ {\bf k}+{\bf Q}}|a_{\bf k}^{\dagger}\rangle\rangle_{\omega}&=&
-\frac{i}{2\pi}
 \frac{\Delta_{\bf k}[(\omega+A_{{\bf k}+})(\omega+A_{{\bf k}-})
      +B_{\bf k}^2-\Delta_{\bf k}^2]}
{\{\omega^2-\omega_{C+}^2({\bf k})\}\{\omega^2-\omega_{C-}^2({\bf k})\}}, \\
\langle\langle a_{-{\bf k}}^{\dagger}|a_{\bf k}^{\dagger}\rangle\rangle_{\omega}&=& 
-\frac{1}{2\pi}
 \frac{B_{\bf k}(\omega^2-A_{{\bf k}+}^2+B_{\bf k}^2-\Delta_{\bf k}^2)}
{\{\omega^2-\omega_{C+}^2({\bf k})\}\{\omega^2-\omega_{C-}^2({\bf k})\}}, \\
\langle\langle a_{-{\bf k}+{\bf Q}}^{\dagger}
                         |a_{\bf k}^{\dagger}\rangle\rangle_{\omega}&=&
+\frac{i}{2\pi}
 \frac{\Delta_{\bf k}B_{\bf k}(A_{{\bf k}+}+A_{{\bf k}-})}
{\{\omega^2-\omega_{C+}^2({\bf k})\}\{\omega^2-\omega_{C-}^2({\bf k})\}}.
\end{eqnarray}
\end{widetext}
They contain complete information about the bosonic 
correlation functions which appear in Eqs. (\ref{szcoa}) and 
(\ref{sscoa}). They are obtained from the temperature Green's functions 
using the fluctuation-dissipation theorem,\cite{Zub60}
\begin{eqnarray}
\label{av1}
%\langle a_i^{\dagger}a_i^{}\rangle&=&
\delta S^z&=&
 \frac{1}{N}\sum_{\bf k}\langle a_{\bf k}^{\dagger}a_{\bf k}^{}\rangle
 \nonumber \\
&=& \frac{1}{N}\sum_{\bf k}\int d\omega
 \frac{2\Im\langle\langle a_{ {\bf k}}|a_{\bf k}^{\dagger}
 \rangle\rangle_{\omega-i\epsilon}}{e^{\beta\omega}-1},    \\
\label{av2}
%\langle a_i^{\dagger}a_j^{\dagger}\rangle&=&
\delta{\cal C}&=&
 \frac{1}{N}\sum_{\bf k}\langle a_{\bf k}^{\dagger}
                      a_{{\bf k}+{\bf Q}}^{}       \rangle 
 \nonumber \\
&=& \frac{1}{N}\sum_{\bf k}\int d\omega
 \frac{2\Im\langle\langle a_{-{\bf k}}^{\dagger}|a_{\bf k}^{\dagger}
 \rangle\rangle_{\omega-i\epsilon}}{e^{\beta\omega}-1},    
\end{eqnarray}
where $\beta=1/k_BT$. 

In a similar way one may find the orbital excitations and the 
respective Green's functions needed to determine the alternation 
of the orbital correlations in the dimerized structure,
\begin{equation}
\langle{\vec\tau}_i\cdot{\vec\tau}_{i+1}\rangle\equiv {\cal T}_{i,i+1}=
{\cal T}_{i,i+1}^{(0)}+\delta{\cal T}e^{i\pi z_i},
\label{ttcoa}
\end{equation}
and the renormalized value of the order parameter,
\begin{equation}
\langle \tau^z_i\rangle\equiv \frac{1}{2}-\delta\tau^z.
\label{tauzcoa}
\end{equation}
As in case of spin operators, we used the rotation (\ref{rotatau})
of orbital operators to the ferro orbital state, followed by the 
Holstein-Primakoff transformation (\ref{ow}) to the respective boson 
operators $\{b_i^{},b_i^{\dagger}\}$. One finds the LOW Hamiltonian, 
\begin{eqnarray}
&H_{\rm LOW}&=
      J\eta(r_1+r_3)\sum_{\langle ij\rangle\parallel ab} 
      (p_i+p_j)  \nonumber \\
\!&+&\!\!\!\!J\!\!\sum_{\langle i,i+1\rangle\parallel c}\!
\left\{\left[R-\frac{1}{2}{\cal C}_{0}[r_1-\eta(r_1+r_3)]\right]
( p_i+p_{i+1} )\right.                         \nonumber \\  
\!&+&\!\!\!\!\left.\left[R-\frac{1}{2}{\cal C}_{0}[r_1-\eta r_1(1-\eta)]\right]
(b_{i+1}^{\dagger}b_{i}^{}+b_{i}^{\dagger}b_{i+1}^{})\right\} \nonumber \\ 
\!&+&\!\!\!\!\frac{1}{2}J{\cal C}_{0}\delta_o r_1(1-\eta)
\sum_{\langle i,i+1\rangle\parallel c}
(b_{i+1}^{\dagger}b_{i}^{}+b_{i}^{\dagger}b_{i+1}^{}).
\label{Hlow}
\end{eqnarray}
where $p_i=b_i^{\dagger}b_i^{}$.

In spite of the 1D nature of orbital dispersion (\ref{owc0}), the 
Fourier transformation to boson operators in the reciprocal 
(momentum) space is three-dimensional and takes here the form
\begin{equation}
b_{k}^{\dagger}=\frac{1}{\sqrt{N}}\sum_i 
                    e^{ ik_zz_i}b_i^{\dagger},   \hskip .3cm
b_{k}^{}       =\frac{1}{\sqrt{N}}\sum_i 
                    e^{-ik_zz_i}b_i^{},
\label{decob}
\end{equation}
Simalr to spin case, we find the energies of orbital wave excitations 
and the average values of the boson operators at finite temperature $T$, 
using temperature Green's functions for boson operators in momentum 
space.\cite{Zub60,Hal72} 
The system of equations is generated by the following equation of motion,
\begin{equation}
\omega\langle\langle b_{\bf k}^{}|b_{\bf k}^{\dagger}\rangle\rangle_{\omega} =
\frac{1}{2\pi}+\langle\langle 
 [b_{\bf k}^{},H_{\rm LOW}]|b_{\bf k}^{\dagger}\rangle\rangle_{\omega},
\label{eqmob}
\end{equation}
and equations for three other Green functions:
$\langle\langle b_{ {\bf k}+{\bf Q}}^{}|b_{\bf k}^{\dagger}\rangle\rangle_{\omega}$,
$\langle\langle b_{-{\bf k}        }^{}|b_{\bf k}^{\dagger}\rangle\rangle_{\omega}$,
and
$\langle\langle b_{-{\bf k}+{\bf Q}}^{}|b_{\bf k}^{\dagger}\rangle\rangle_{\omega}$,
follow. The following expressions 
define the algebraic structure of the orbital problem:
\begin{eqnarray}
\label{abdorb}
\bar{A}_{\bf k}&=&r_1+\eta(r_1+r_3)-\frac12 (1-{\cal C}_0)[r_1-\eta(r_1+r_3)]  
\nonumber \\
     &+&\frac12 (1+2\eta r_1-\eta r_3) Y_{ab},                            \\
\bar{B}_{\bf k}&=&\Big[r_1-\frac12 (1-{\cal C}_0)r_1\big(1-\eta\big)\Big],             \\
\Theta_{\bf k}&=&\frac12 (1-{\cal C}_0)\delta_or_1\big(1-\eta\big) \sin k_z.
\end{eqnarray}
Here we used the short hand notation for the spin correlation function
in $ab$ planes, 
\begin{equation}
Y_{ab}\equiv\langle{\vec S}_i\cdot{\vec S}_j\rangle+1,
\label{yab}
\end{equation}
for a bond $\langle ij\rangle\parallel ab$.

The respective system of equations of motion has a similar structure
to that of Eq. 
(\ref{eqmo}) is:
\begin{widetext}
\begin{equation}
\label{gfeqo}
\left(\begin{array}{cccc}
 \bar{A}_{{\bf k}}-\Omega_C({\bf k}) & i\Theta_{\bf k} & \bar{B}_{\bf k} & 0     \\
-i\Theta_{\bf k} & \bar{A}_{{\bf k}}-\Omega_C({\bf k}) & 0 & -\bar{B}_{\bf k}    \\
-\bar{B}_{\bf k} & 0 &  -\bar{A}_{{\bf k}}-\Omega_C({\bf k}) & -i\Theta_{\bf k}  \\
 0 &  \bar{B}_{\bf k} & i\Theta_{\bf k} &   -\bar{A}_{{\bf k}}-\Omega_C({\bf k})
\end{array} \right)
\left( \begin{array}{c}
 \langle\langle b_{ {\bf k}}          |b_{\bf k}^{\dagger}\rangle\rangle_{\omega} \\
 \langle\langle b_{ {\bf k}+{\bf Q}}  |b_{\bf k}^{\dagger}\rangle\rangle_{\omega} \\
 \langle\langle b_{-{\bf k}}^{\dagger}|b_{\bf k}^{\dagger}\rangle\rangle_{\omega} \\
 \langle\langle b_{-{\bf k}+{\bf Q}}^{\dagger}
                                      |b_{\bf k}^{\dagger}\rangle\rangle_{\omega}
\end{array} \right) = 
-\frac{1}{2\pi}\left( \begin{array}{c}
 1   \\
 0   \\
 0   \\
 0
\end{array} \right).
\end{equation}
\end{widetext}
The Green functions can be now found from Eqs. (\ref{gfeqo}).
They contain complete information about the bosonic correlation 
functions which appear in Eqs. (\ref{ttcoa}) and (\ref{tauzcoa}). 
They are obtained from the temperature Green's functions 
using the fluctuation-dissipation theorem,\cite{Zub60}
\begin{eqnarray}
\label{at1}
%\langle a_i^{\dagger}a_i^{}\rangle&=&
\delta \tau^z&=&
 \frac{1}{N}\sum_{\bf k}\langle b_{\bf k}^{\dagger}b_{\bf k}^{}\rangle
 \nonumber \\
&=& \frac{1}{N}\sum_{\bf k}\int d\omega
 \frac{2\Im\langle\langle b_{ {\bf k}}|b_{\bf k}^{\dagger}
 \rangle\rangle_{\omega-i\epsilon}}{e^{\beta\omega}-1},    \\
\label{at2}
%\langle a_i^{\dagger}a_j^{\dagger}\rangle&=&
\delta{\cal T}&=&
 \frac{1}{N}\sum_{\bf k}\langle b_{\bf k}^{\dagger}
                      b_{{\bf k}+{\bf Q}}^{}       \rangle 
 \nonumber \\
&=& \frac{1}{N}\sum_{\bf k}\int d\omega
 \frac{2\Im\langle\langle b_{-{\bf k}}^{\dagger}|b_{\bf k}^{\dagger}
 \rangle\rangle_{\omega-i\epsilon}}{e^{\beta\omega}-1}.    
\end{eqnarray}
The values of the orbital correlation functions Eqs. (\ref{ttcoa}) and 
(\ref{tauzcoa}) in the dimerized structure were used together with the 
respective spin correlation functions Eqs. (\ref{szcoa}) and 
(\ref{szcoa}) to obtain the self-consistent solution of Fig. 
\ref{fig:dios}.

%%%%%%%%%%%%%%%%%%%%%%%%%%%%%%%%%%%%%%%%%%%%%%%%%%%%%%%%%%%%%%%%%%%%%%%%
%%
%%                           REFERENCES
%%
%%%%%%%%%%%%%%%%%%%%%%%%%%%%%%%%%%%%%%%%%%%%%%%%%%%%%%%%%%%%%%%%%%%%%%%%


\begin{thebibliography}{00}
\bibitem{notex} The excitation energies to $t_{2g}^3t_{2g}^1$ 
                configuration on a bond $\langle ij\rangle$ expressed 
		in terms of the Racah parameters are: $A-5B$ ($^4A_2$), 
                $A+4B+3C$ ($^2E$ and $^2T_1$), and $A+10B+5C$ ($^2T_2$); 
                this spectrum is rigorously reproduced with $U$ and $J_H$ 
		given by Eqs. (\ref{uj}), see Ref. \onlinecite{Gri71}.
\bibitem{noteo} The off-diagonal elements of the Coulomb interaction
                $\propto J_H$ are also responsible for the propagation
                of longitudinal orbital excitations in $e_g$ systems,
                as explained in Ref. \protect\onlinecite{Ole00}.
\bibitem{noteowg} This correction could easily be included but it does 
                  not modify the qualitative conclusions --- 
                  this energy contribution would only slightly modify 
		  the critical value of $V$ in the phase diagram of 
		  Fig. \ref{fig:phd}.
\end{thebibliography}
\end{document}